\documentclass[iop,twocolappendix,numberedappendix]{emulateapj}

\def\simlt{\mathrel{\rlap{\lower 3pt\hbox{$\sim$}}\raise 2.0pt\hbox{$<$}}}
\def\simgt{\mathrel{\rlap{\lower 3pt\hbox{$\sim$}} \raise 2.0pt\hbox{$>$}}}

\def\gtsima{$\; \buildrel > \over \sim \;$}
\def\ltsima{$\; \buildrel < \over \sim \;$}
\def\gtrsim{\lower.5ex\hbox{\gtsima}}
\def\lesssim{\lower.5ex\hbox{\ltsima}}

\usepackage{natbib}

\newcommand{\q}{\begin{equation}}
\newcommand{\qa}{\begin{eqnarray}}
\newcommand{\qs}{\begin{eqnarray*}}
\newcommand{\nq}{\end{equation}}
\newcommand{\nqa}{\end{eqnarray}}
\newcommand{\nqs}{\end{eqnarray*}}

\shorttitle{{\it In situ} formation of SgrA$^\ast{}$ stars}
\shortauthors{Mapelli et al.}

\begin{document}

\title{{\it In situ} formation of SgrA$^\ast{}$ stars via disk fragmentation: parent cloud properties and thermodynamics}
\author{M. Mapelli\altaffilmark{1}, T. Hayfield\altaffilmark{2}, L. Mayer\altaffilmark{3}, J. Wadsley\altaffilmark{4}}

\altaffiltext{1}{INAF-Osservatorio astronomico di Padova, Vicolo dell'Osservatorio 5, I--35122, Padova, Italy; {\tt michela.mapelli@oapd.inaf.it}}
\altaffiltext{2}{Max-Planck-Institut f\"ur Astronomie, K\"onigstuhl 17, D--69117, Heidelberg, Germany}
\altaffiltext{3}{Institute for Theoretical Physics, University of Z\"urich, Winterthurerstrasse 190, CH-8057, Z\"urich, Switzerland}
\altaffiltext{4}{Department of Physics and Astronomy, McMaster University, Hamilton, ON L8S 4M1, Canada}


\begin{abstract}
The formation of the massive young stars surrounding  SgrA$^\ast{}$ is still an open question. In this paper, we simulate the infall of a turbulent molecular cloud towards the Galactic Center (GC). We adopt two different cloud masses ($4.3\times{}10^4$ M$_\odot{}$ and $1.3\times{}10^5$ M$_\odot{}$). We  run five simulations: the gas is assumed to be isothermal in four runs, whereas radiative cooling is included in the fifth run.
In all the simulations, the molecular cloud is tidally disrupted, spirals towards the GC, and forms a small, dense and eccentric disk around SgrA$^\ast{}$.
With high resolution simulations, we follow the fragmentation of the gaseous disk. Star candidates form in a ring at $\sim{}0.1-0.4$ pc from the super-massive black hole (SMBH) and have moderately eccentric orbits ($e\sim{}0.2-0.4$), in good agreement with the observations. 
The mass function of star candidates is top-heavy only if the local gas temperature is high ($\gtrsim{}100$ K) during the star formation and if the parent cloud is sufficiently massive ($\gtrsim{}10^5$ M$_\odot{}$).
Thus, this study indicates that the infall of a massive molecular cloud is a viable scenario for the formation of massive stars around SgrA$^\ast{}$, provided that the gas temperature is kept sufficiently high ($\gtrsim{}100$ K). 
\end{abstract}
\keywords{Methods: numerical - Galaxy : center - Stars: formation - ISM: clouds}

\section{Introduction}
The origin of young massive stars that crowd the Galactic Center (GC) has been a puzzle for a long time. Most of the massive stars observed in the central parsec reside in one or
perhaps two disks: a well-defined clockwise-rotating disk, and a counter-clockwise disk, whose existence is still debated (Genzel et al. 2003, hereafter G03; Paumard et al. 2006, hereafter P06; Lu et al. 2009; Bartko et al. 2009, 2010; see Genzel et al. 2010 for a recent review).
These disks have well-defined inner ($r_{in}\sim{}0.04$ pc) and outer radii ($r_{out}\sim{}0.5$ pc). 
The $S$ stars observed at distances $\lesssim{}0.02$ pc from SgrA$^\ast{}$, the source identified with the super massive black hole (SMBH), have randomly oriented motions and do not belong to the disks (G03; Ghez et al. 2005; Eisenhauer et al. 2005).
The massive stars inside the disks are young ($6\pm{}2$ Myr, P06) and must have formed over a short period ($<2$ Myr, P06). Their estimated initial mass function (IMF) is heavier than Salpeter's one (P06; Bartko et al. 2009). The total mass in the disks cannot exceed $1.5\times{}10^4$ M$_\odot{}$, but is more likely of the order of $5\times{}10^3$ M$_\odot{}$ (P06).

Such stars cannot have formed {\it in situ}
in `normal' conditions, as the tidal forces exerted from the SMBH would have disrupted
 the parent molecular cloud (Levin \& Beloborodov 2003; G03). Thus, an alternative scenario has been proposed, according to which a young cluster spiraled towards the GC and deposited its stars around SgrA$^\ast$ (Gerhard 2001; McMillan \& Portegies Zwart 2003; Portegies Zwart, McMillan \& Gerhard 2003; 
Kim \& Morris 2003; Kim, Figer \&{}  Morris 2004; 
G\"urkan \& Rasio 2005; Fujii et al. 2008, 2009).  However, even the latter scenario suffers from various shortcomings, such as the premature disruption of the cluster and the excessively 
long dynamical friction time. These problems have only been partially solved by assuming that the original clusters 
host intermediate-mass black holes (Portegies Zwart et al. 2006), or by using new computational schemes (Fujii et al. 2008).

On the other hand, the problem of tidal forces exerted by the SMBH can be overcome if,
 at some point 
in the past, 
a dense gaseous disk existed around SgrA$^\ast{}$. Such a disk could have formed because of the infall and tidal disruption of a molecular cloud. If the density in the disk was high enough, 
it might have become unstable to fragmentation and formed stars (Poliachenko \&{} Shukhman 1977; Kolykhalov \&{} Sunyaev 1980; Shlosman \&{} Begelman 1987; Sanders 1998; Levin \& Beloborodov 2003; G03; Goodman 2003; Milosavljevic \& Loeb 2004; Nayakshin \& Cuadra 2005; Rice et al. 2005; Alexander et al. 2008; Collin \& Zahn 2008; Wardle \&{} Yusef-Zadeh 2008; Yusef-Zadeh \&{} Wardle 2011). The (nearly) absence of massive stars at distances $>0.5$ pc from SgrA$^\ast{}$ supports this idea of {\it in situ} formation (Nayakshin \& Sunyaev 2005; P06).  This scenario is also favored by the existence of two massive molecular clouds within $\sim{}20$ pc from the dynamical center of our Galaxy (Solomon et al. 1972). On-going star formation (SF) is associated with one of these two clouds, named M$-0.02-0.07$ (Yusef-Zadeh et al. 2010). The other cloud (M$-0.13-0.08$) is highly elongated toward SgrA$^\ast{}$ and has a `finger-like' extension pointing in the direction of the circumnuclear ring (Okumura et al. 1991; Ho et al. 1991; Novak et al. 2000, and references therein). The circumnuclear ring of dense molecular gas lies at a projected distance of $2-7$ pc from the GC (e.g., Genzel et al. 1985; G\"usten et al. 1987; Scoville et al. 2003; Christopher et al. 2005) and is on the verge of forming stars (Yusef-Zadeh et a. 2008; 2011). Furthermore, the innermost 3 pc of the GC are extremely rich of ionized gas, organized in at least three streams (e.g., Scoville et al. 2003; Zhao et al. 2009, and references therein). Recently, a small ($\sim{}3$ Earth masses) and relatively cold cloud (gas temperature $\approx{}10^4$ K)  has been observed on its way towards the SMBH, with a orbital plane coincident with that of the clockwise stellar disk (Gillessen et al. 2012; Burkert et al. 2012).

Nayakshin, Cuadra \& Springel (2007, hereafter NCS07) and Alexander et al. (2008) simulated SF in a gaseous disk around SgrA$^\ast{}$, and found encouraging results for this scenario. However, NCS07 and Alexander et al. (2008) assume that the gaseous disk was already in place when it started forming stars, and do not consider the process that lead to the formation of the disk itself. Bonnell \&{} Rice (2008, hereafter BR08), Mapelli et al. (2008, hereafter M08) and Alig et al. (2011, hereafter A11) simulate the infall of a molecular cloud toward SgrA$^\ast{}$ and study the formation of a dense gaseous disk around the SMBH.  Hobbs \&{} Nayakshin (2009, hereafter HN09) simulate the collision of two molecular clouds leading to the formation of one or more disks around the SMBH. A11 study the density distribution of the gas in the circumnuclear disk, but do not trace the evolution of the system up to the fragmentation and to the formation of the proto-stars. BR08, M08 and HN09 use the sink-particle technique to model SF in the gaseous disk around the SMBH. This was required by the huge density range between the `background' gas and the particles that are collapsing into stars. However, the sink-particle technique might be affected by various uncertainties, because it does not trace the fragmentation of the cloud, but it replaces this process with the formation of sink particles.
In this paper, we simulate the infall of a molecular cloud toward SgrA$^\ast{}$ and we study the ongoing SF around the SMBH without adopting the sink-particle method. Our results indicate that the possibility of forming massive stars around  SgrA$^\ast{}$ strongly depends on the initial mass of the infalling cloud and on the temperatures that are locally reached by the gas.

\section{Models and simulations}
We ran N-body/Smoothed Particle Hydrodynamics (SPH) simulations of a molecular cloud evolving in a potential dominated by the SMBH.
We used the SPH code GASOLINE (Wadsley, Stadel \&{} Quinn 2004), upgraded with the Read et al. (2010) OSPH modifications, to address the SPH limitations outlined, most recently, by Agertz et al. (2007).

In the simulations, the SMBH is represented by a sink particle, with initial mass $M_{\rm BH}=3.5\times{}10^6$ M$_\odot{}$ (Ghez et al. 2003), sink radius $r_{acc}=5\times{}10^{-3}$ pc and softening radius $\epsilon{}=1\times{}10^{-3}$ pc. We also add a rigid potential, according to a density distribution $\rho{}(r)=\rho{}_0\,{}(r/c)^{-\alpha{}}$, where $\rho{}_0=1.2\times{}10^6$ M$_\odot\textrm{ pc}^{-3}$, $c=0.39$ pc, and $\alpha{}=1.4$ at $r<c$ and $=2.0$ at $r>c$ (G03; although recent papers indicate a more complex scenario, e.g., Yusef-Zadeh, Bushouse \&{} Wardle 2012).  Before this paper, the cuspy rigid potential was introduced only by M08 and by HN09, whereas other papers (e.g., BR08; A11) consider the SMBH and the cloud in isolation.

The clouds used in this experiment are spherical with a radius of 15 pc, 
are seeded with supersonic turbulent velocities and marginally self-bound.
 To simulate interstellar
turbulence, the velocity field of the cloud was generated on a grid as a divergence-free
Gaussian random field with an imposed power spectrum P($k$), varying as $k^{-4}$.
This
yields a velocity dispersion $\sigma{}(l)$, varying as $l^{1/2}$, chosen to agree with
the Larson scaling relations (Larson 1981). The velocities were then interpolated from the
grid to the particles. 

 \begin{deluxetable}{lllll}
 \tabletypesize{\tiny}
 \tablewidth{0pt}
 \tablecaption{Initial conditions.}
\tablehead{\colhead{Run}
& \colhead{$N_{\rm p}/10^6$}
& \colhead{M$_{\rm MC}/{\rm M}_\odot{}$}
& \colhead{T$_{\rm MC}$ [K]}
& \colhead{E.O.S.}}
\startdata
A       & 3.221398       & $1.3\times{}10^5$  & 500 & isothermal\\
B       & 3.221398       & $1.3\times{}10^5$  & 100 & isothermal \\
C       & 1.073799       & $4.3\times{}10^4$  & 500 & isothermal\\
D       & 1.073799       & $4.3\times{}10^4$  & 100 & isothermal \\
E       & 3.221398       & $1.3\times{}10^5$  & 100 & cooling 
\enddata
\tablecomments{
{\it Notes.} $N_{\rm p}$ is the number of gas particles in each run, M$_{\rm MC}$ and T$_{\rm MC}$ are the total mass and the initial temperature of the simulated molecular cloud, respectively. E.O.S. indicates the equation of state of gas.}
\end{deluxetable}

\begin{figure*}
\center{{
\includegraphics[height=14cm]{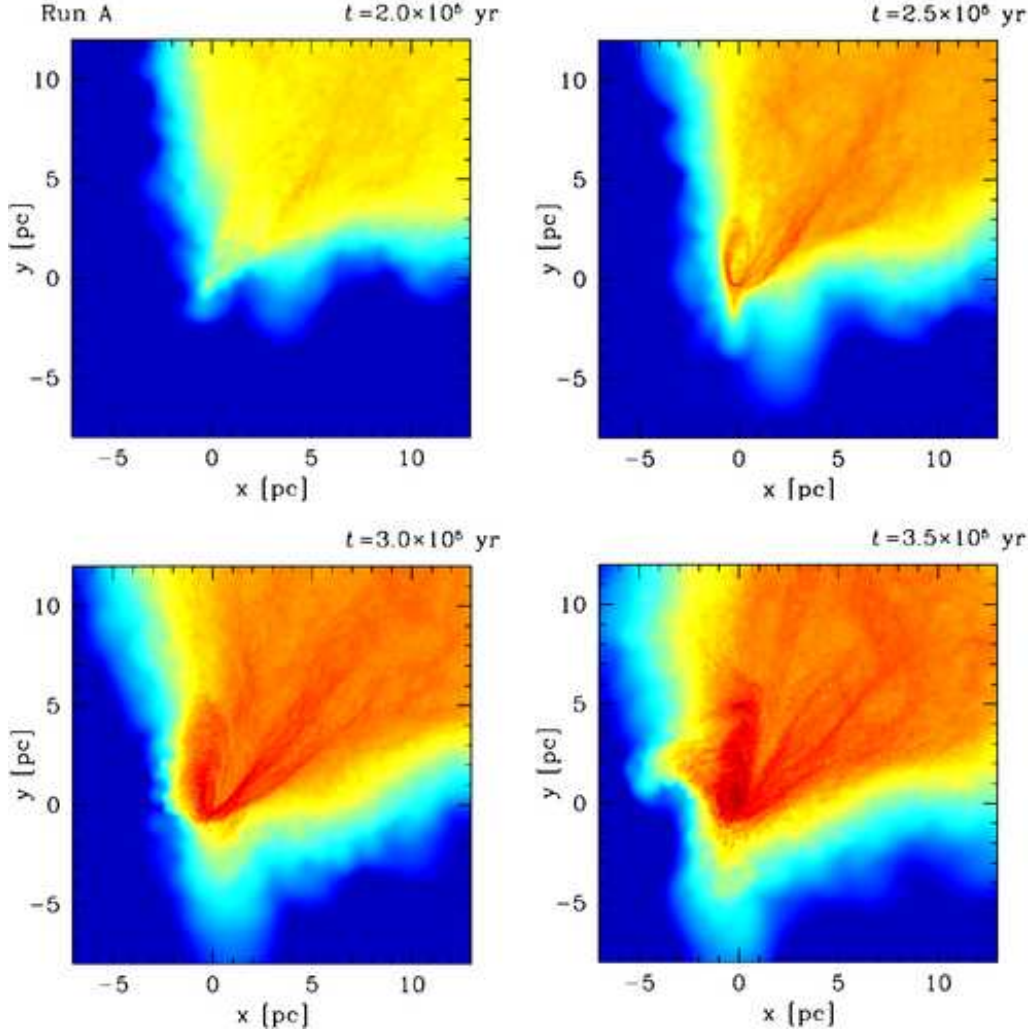} 
}}
\caption{\label{fig:fig1} Density map of gas in the cloud (run~A), projected along the $z-$axis. The frames measure 20 pc per edge and are centered at (x, y, z)=(3, 2, 0) pc from the GC. The density ranges from $2.23\times{}10^{-2}$ to $4.45\times{}10^3$ M$_\odot{}$ pc$^{-2}$ in logarithmic scale. From left to right and from top to bottom: $t=2.0\times{}10^5$ yr, $2.5\times{}10^5$ yr, $3.0\times{}10^5$ yr and $3.5\times{}10^5$ yr. 
}
\end{figure*}
\begin{figure*}
\center{{
\includegraphics[height=4.0cm]{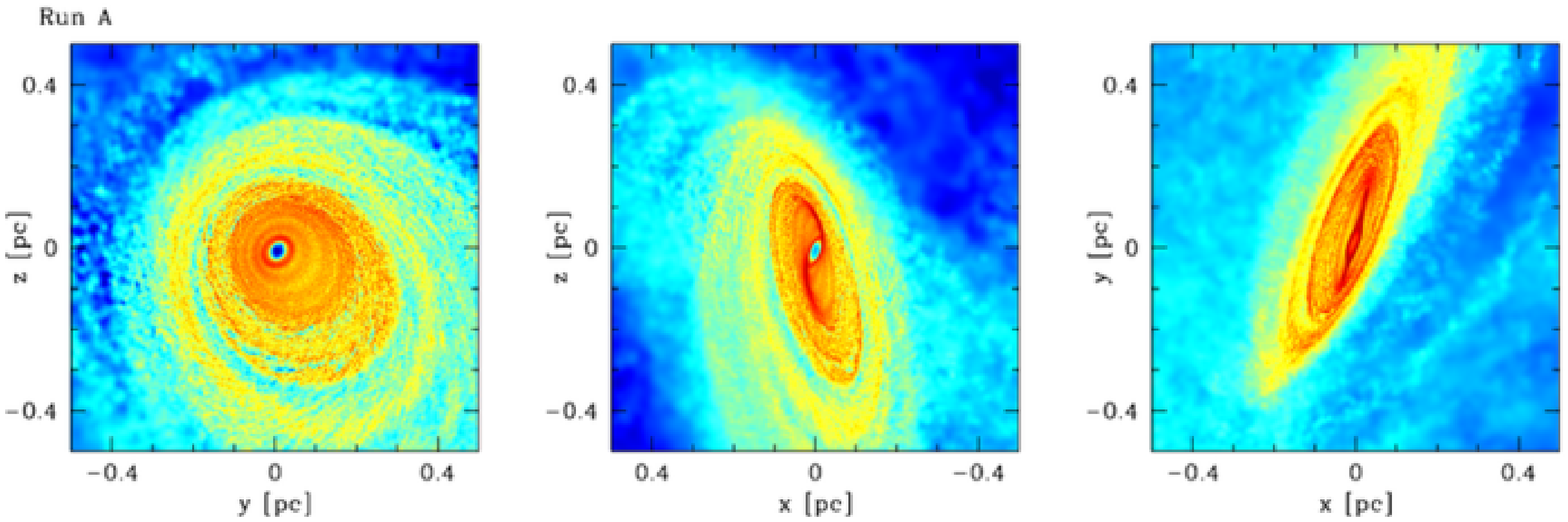}  
\includegraphics[height=4.0cm]{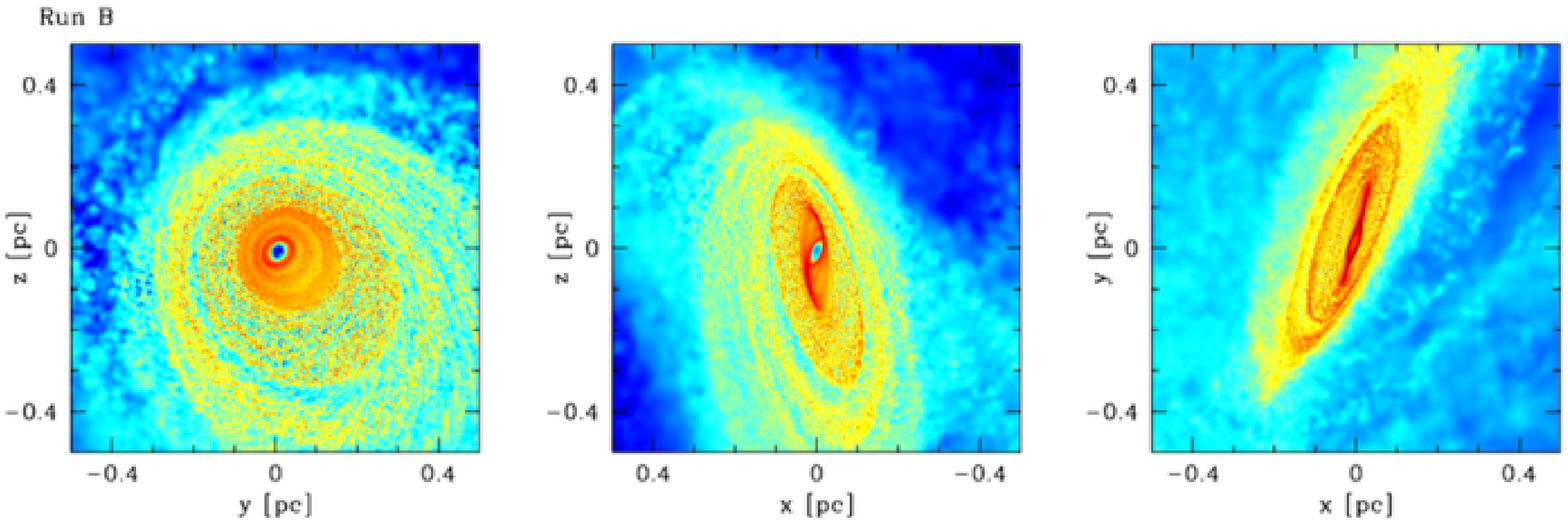}  
\includegraphics[height=4.0cm]{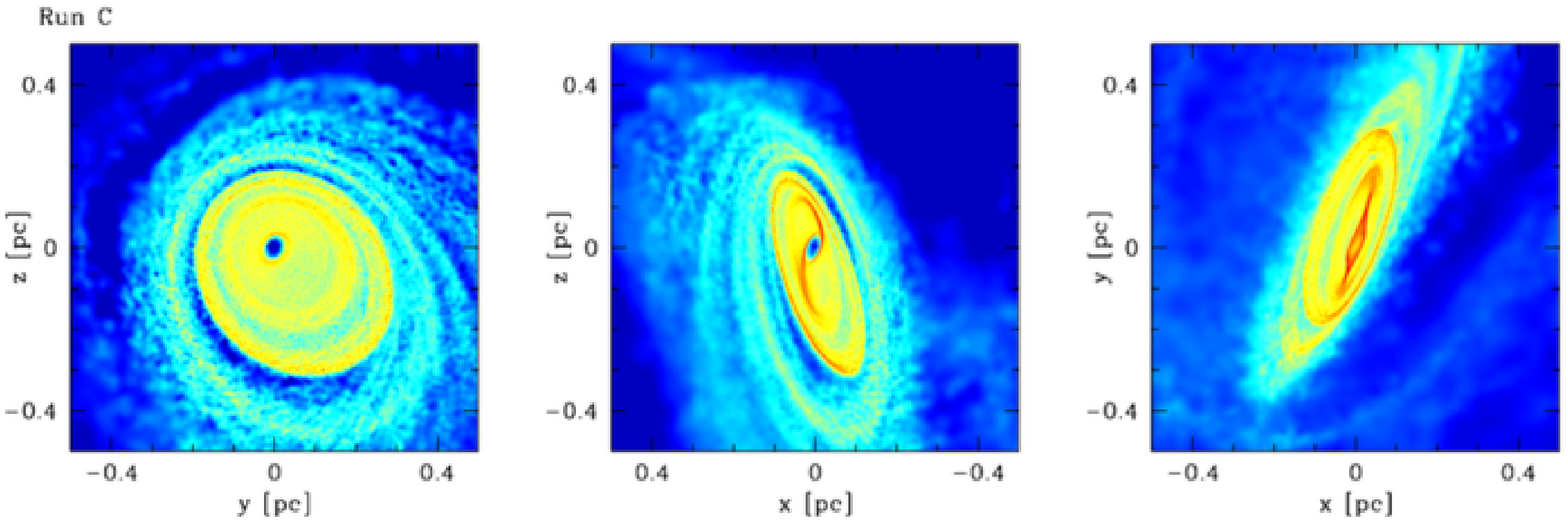}  
\includegraphics[height=4.0cm]{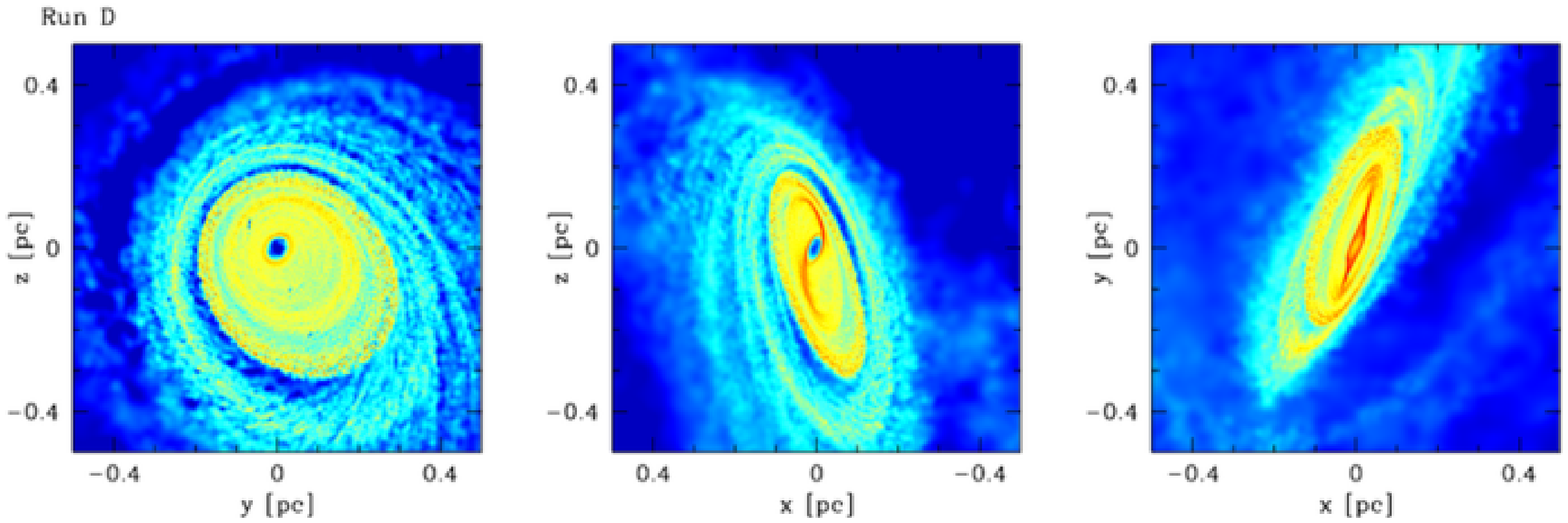}  
\includegraphics[height=4.0cm]{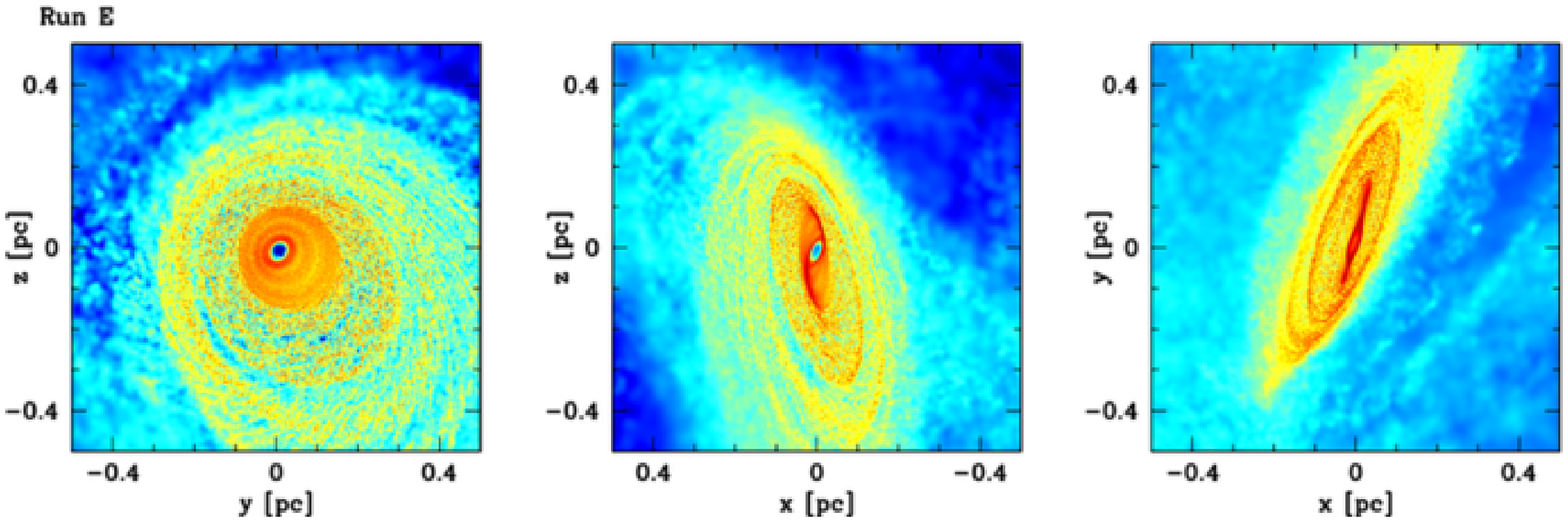}  
}}
\caption{\label{fig:fig2} Density map of gas in the central disk, projected along the $x-$ (left-hand panels), $y-$ (central panels) and $z-$axis (right-hand panels) at $t=4.8\times{}10^5$ yr. From top to bottom: run~A, B, C, D and E. The frames measure 1 pc per edge and are centered in the GC. The density ranges from $7.05\times{}10^{2}$ to $7.05\times{}10^5$ M$_\odot{}$ pc$^{-2}$ in logarithmic scale. %
}
\end{figure*}

\begin{figure*}
\center{{
\includegraphics[height=14cm]{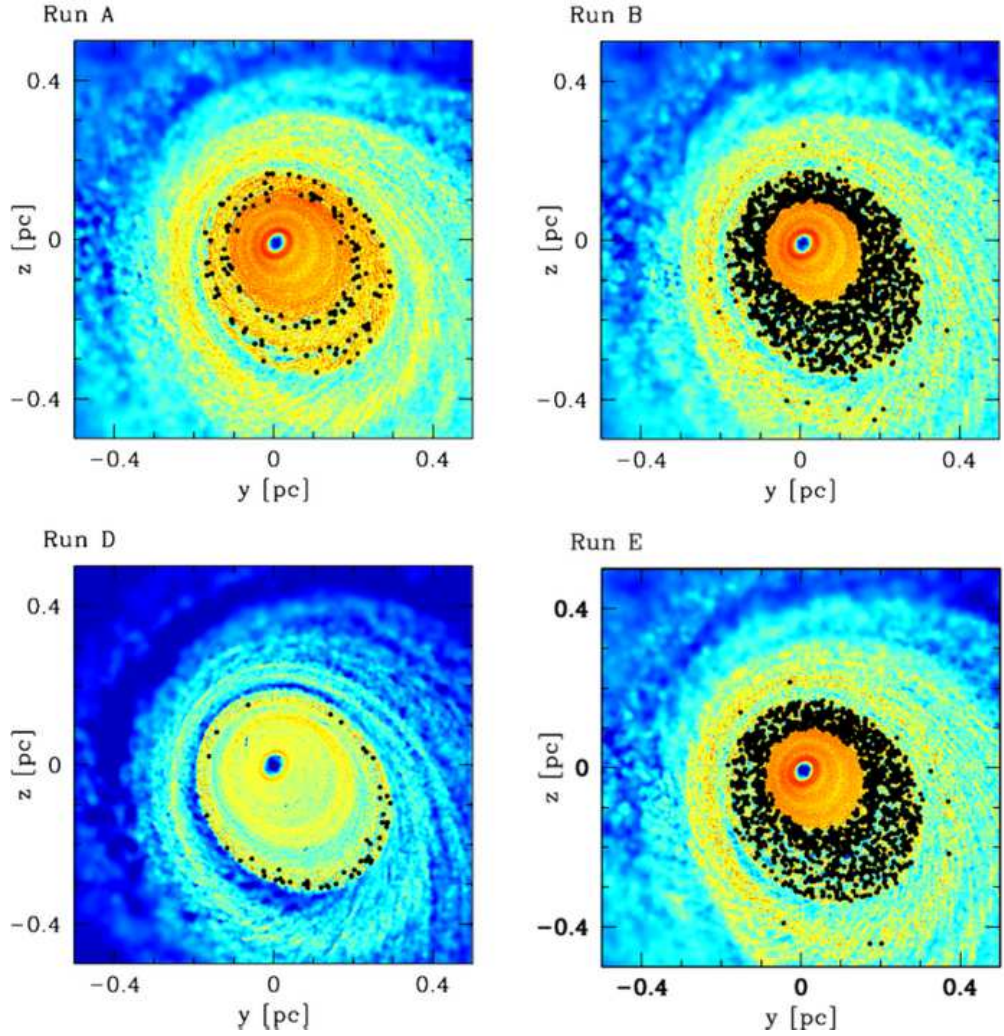} 
}}
\caption{\label{fig:fig3} 
 Density map of gas in the central disk, projected along the $x-$axis at $t=4.8\times{}10^5$ yr. From top to bottom and from left to right: run~A, B, D and E. Run~C is not shown because no stars have formed. The frames measure 1 pc per edge and are centered in the GC. The density ranges from $7.05\times{}10^{2}$ to $7.05\times{}10^5$ M$_\odot{}$ pc$^{-2}$ in logarithmic scale.
The superimposed filled black circles indicate the star candidates, i.e. the positions where stars are expected to form (see the text for details). 
}
\end{figure*}
\begin{figure}
\center{{
\includegraphics[height=8.5cm]{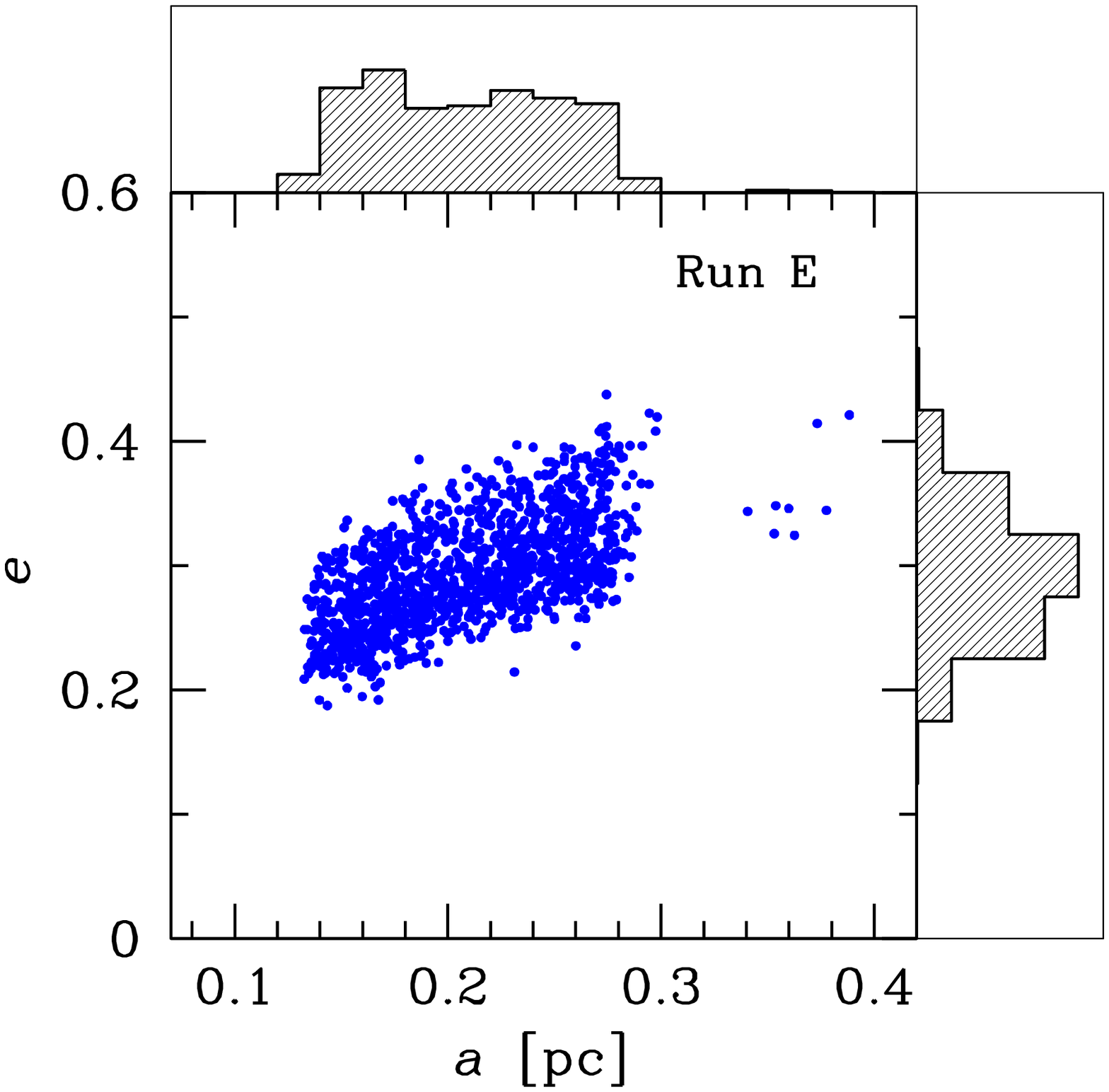} 
}}
\caption{\label{fig:orbits} 
Eccentricity $e$ versus semi-major axis $a$ at $t=4.8\times{}10^5$ yr in run~E. The marginal histograms show the distribution of $a$ (top histogram) and $e$ (right-hand histogram). 
}
\end{figure}
The center-of-mass of the cloud is initially at 25 pc from the SMBH. The orbit of the cloud was chosen so that the impact parameter\footnote{The adopted impact parameter is very small, to guarantee a low orbital angular momentum and to increase the chance that the core of the parent cloud engulfs the SMBH (Wardle \&{} Yusef-Zadeh 2008). This choice is biased, as it likely favors the formation a dense gas disk. The properties of the resulting gas disk are expected to depend on the impact parameter, as well as on the other (both orbital and intrinsic) properties of the cloud. On the other hand, BR08 adopt a 10 times larger value for the impact parameter (0.1 pc) than in this paper, and the differences in the properties of the gaseous disk (e.g., scale-length, thickness, outer and inner radius) are not significant. The role of the impact parameter as well as that of  the other orbital properties deserve further studies.} with respect to the SMBH is $10^{-2}$ pc and the initial velocity is close to the escape velocity from the SMBH at the initial distance (i.e. the orbit is marginally bound and highly eccentric). As the tidal density  at $\sim{}25$ pc from the GC is higher than the average initial density of our cloud, our marginally bound cloud must have formed further out and then migrated closer to the GC. Various processes could have brought the cloud in such position. For example, the cloud might have achieved this orbit after a collision with another cloud (e.g., Wardle \&{} Yusef-Zadeh 2008; HN09; A11). On the other hand, the existence of two giant molecular clouds, M$-0.02-0.07$ and M$-0.13-0.08$, at $\sim{}7$ and $\sim{}13$ pc, respectively, from the GC (Solomon et al. 1972; Okumura et al. 1991; Ho et al. 1991; Novak et al. 2000) shows that molecular clouds, probably unbound, exist at $\lesssim{}20$ pc from the GC. 

The cloud is assumed to be isothermal in runs A$-$D. This assumption neglects the local variations of the effective equation of state, that might occur since the balance between heating and cooling processes depends on local conditions (see Section~3.3).  In run~E, we include radiative cooling. The radiative cooling algorithm is the same as that described in Boley (2009) and in Boley et al. (2010). The cooling is calculated from $\nabla{}\cdot{}F=-(36\,{}\pi{})^{1/3}\,{}s^{-1}\sigma{}\left({\rm T}^4-{\rm T}^4_{\rm irr}\right)\,{}(\Delta{}\tau{}+1/\Delta{}\tau{})^{-1}$, where $s=\left(m/\rho{}\right)^{1/3}$ and $\Delta{}\tau{}=s\,{}\kappa{}\,{}\rho{}$, for the local opacity $\kappa{}$, particle mass $m$ and density $\rho{}$. D'Alessio et al. (2001) opacities are used, with a 1 $\mu{}$m maximum grain size. The irradiation temperature is T$_{\rm irr}=100$ K everywhere.

In this paper, we run five simulations, with two different masses for the cloud ($1.3\times{}10^5$ and $4.3\times{}10^4$ M$_\odot{}$) and two different initial gas temperatures (100 and 500~K). The initial temperature of the gas is set to be quite high for a Galactic molecular cloud, as observational data (see e.g. Nagai et al. 2007; Oka et al. 2007) suggest that the gas temperature is $\gtrsim{}100$~K in the vicinity of Sgr~A$^\ast{}$. The initial masses of the simulated clouds, the initial temperature of the gas and the adopted equation of state in the various runs are listed in Table~1. 
In all the runs, the mass of the gas particles is 0.04~M$_\odot{}$ and the softening length $10^{-3}$~pc. 

\section{Results}
\subsection{Evolution of the gaseous disk}
In all the runs, the cloud, orbiting around SgrA$^\ast{}$,  rapidly ($t\lesssim{}10^5$ yr) stretches towards the SMBH and is partially disrupted. The branch of the cloud that points toward the SMBH 
begins to spiral in 
towards the GC (see Fig.~\ref{fig:fig1}, in the case of run~A).
A dense, rotating gaseous disk forms at $t\sim{}3-3.5\times{}10^5$ yr at the location of the in-spiraling branch. 
In runs A, B  and E (C and D), the initial density\footnote{Here and in the following, we assume molecular weight $\mu{}=2.46$, the same as in our simulations, consistently with Galactic molecular clouds.} of the gaseous disk is $\sim{}2-8\times{}10^4$ cm$^{-3}$ ($\sim{}1-5\times{}10^4$ cm$^{-3}$) and its initial mass is $\sim{}1100-4000\,{}{\rm M}_\odot{}$ ($\sim{}330-1230\,{}{\rm M}_\odot{}$). The outer radius of the disk is $r_{\rm out}\sim{}0.5$ pc.
 At this stage, the parent cloud is still feeding it through a finger-like structure (see Fig.~\ref{fig:fig1}). 

The gaseous disk
is not  homogeneous or regular, but it is the assembly of many concentric annuli, that spiral around the SMBH, with slightly different inclination and thickness (see Fig.~\ref{fig:fig2}). However, all the structures that form inside $\approx{}2$ pc lie almost on the same plane as the disk: we do not see the formation of any relevant structure with significant inclination ($>>10^{\circ}$) in the very central region.

The initial average thickness of the disk, defined as the ratio between the average scaleheight $h$ and $r_{\rm out}$, is $\zeta{}\sim{}0.1$. 

At $t\sim{}4.8\times{}10^5$ yr the disk has the average density of $\sim{}4\times{}10^5$ cm$^{-3}$ in runs A, B and E, and of $\sim{}2\times{}10^5$ cm$^{-3}$ in runs C and D (see Fig.~\ref{fig:fig2}). However, 
local densities $\gg{}10^{9}$ cm$^{-3}$ are 
reached (see the density map of gas in Fig.~\ref{fig:fig2}). At this stage, the total mass of the gaseous disk is $\sim{}2.1\times{}10^4\,{}{\rm M}_\odot{}$ in runs A, B and E, and $\sim{}7.1\times{}10^3\,{}{\rm M}_\odot{}$ in runs C and D. 

The outer radius  and the thickness of the disk are still $r_{\rm out}\sim{}0.5$ pc and $\zeta{}\sim{}0.1$: they do not change during the entire simulation.  The size of the gaseous disk agrees with the estimate by Wardle \&{} Yusef-Zadeh (2008, 2011), based on the Bondi-Hoyle-Lyttleton formalism (Hoyle \&{} Lyttleton 1939) for the capture radius of a black hole. Applying this formalism to our simulated clouds, we predict a disk radius $r_{\rm out}\sim{}0.55\,{}\lambda^2_{0.3}\,{}(M_{\rm BH}/3.5\times{}10^6 {\rm M}_\odot{})\,{}(v/100\,{}{\rm km\,{}s}^{-1})^{-2}$ pc (where $v$ is the velocity of the cloud when entering the influence radius of the SMBH and $\lambda_{0.3}=0.3$ is the average ratio of the specific angular momentum of a fluid element settled in the final disk to its initial angular momentum, see equation 5 of Wardle \&{} Yusef-Zadeh 2011 for details).
The disk appears distorted at the edges, where fresh gas is being fed by the parent cloud.
The orbits of gas particles have eccentricity in the range $0\le{}e\le{}0.6$.

\subsection{Formation of the stars}
In our analysis, we assume that a star (or, more precisely, a stellar candidate) is formed, when a spherical clump of $\ge{}32$ particles (where 32 is the number of neighbors in our code) and radius $r_\ast{}\le{}r_{\rm th}=2.2\times{}10^{-3}\,{}({\rm T}_{\rm MC}/500)^{1/2}$~pc reaches an average density $\rho{}_\ast{}\ge{}\rho{}_{\rm th}=2\times{}10^{12}$ cm$^{-3}$ (assuming molecular weight $\mu{}=2.46$), i.e. it has an average density $\ge{}10^7$ times higher than the average density of the disk. Under such conditions, the spherical clump is gravitationally bound and completely decoupled from the surrounding particles. The threshold values $r_{\rm th}$ and $\rho{}_{\rm th}$ have been selected on the basis of the density maps (see Fig.~\ref{fig:fig2}). We notice that the threshold value $\rho{}_{\rm th}=2\times{}10^{12}$ cm$^{-3}$ is well above the maximum tidal density inside the simulated disk, which is $\rho_{\rm tid}\le{}10^9$ cm$^{-3}$. $r_{\rm th}\propto{}{\rm T}_{\rm MC}^{1/2}$ accounts for the behavior of the Jeans instability (Jeans 1919). There are no significant changes in the stellar mass function (MF) for $\epsilon{}\le{}r_{\rm th}\le{}4\times{}10^{-3}\,{}({\rm T}_{\rm MC}/500)^{1/2}\,{}{\rm pc}$, where  $\epsilon{}=10^{-3}{\rm pc}$ is the softening, and for $10^{11}\le{}\rho{}_{\rm th}/{\rm cm}^{-3}\le{}10^{13}$.
Furthermore, our choice of the threshold values is quite conservative, as it allows to exclude marginally bound objects. For further details about the choice of $\rho{}_{\rm th}$ and $r_{\rm th}$, see the discussion in Appendix~A.

\subsubsection{The stellar disk}
The first gravitationally bound clumps, i.e. the first stellar candidates form at $t\sim{}4.5\times{}10^5$ yr. We continue the integration up to $t=4.8\times{}10^5$ yr, in order to account for possible mass accretion of the proto-star and to allow most of gravitational instabilities in the disk to evolve. For $t>4.8\times{}10^5$ yr the integration slows down dramatically in run~A,~B and E, because of the large number of collapsed clumps. 

In Fig.~\ref{fig:fig3} the star candidates are marked by filled circles, superimposed to the density map. Run~C is the only simulation in which no stars have formed yet at $t=4.8\times{}10^5$ yr. In run~A, B, D and E the star candidates are distributed in a ring, corotating with the gaseous disk. The stellar ring has a thickness $\zeta{}\sim{}0.05-0.08$. In run~A we can almost distinguish two concentric rings.  In runs~A, B, D and E the ring (and therefore the orbits of star candidates) is eccentric, with minimum (maximum) eccentricity $e\sim{}0.19$ ($e\sim{}0.44$). The average eccentricity is $\langle{}e\rangle{}=0.29\pm{}0.04$. Fig.~\ref{fig:orbits} shows the eccentricity $e$ versus the semi-major axis $a$ for all the stars in run~E. The eccentricity and the semi-major axis were derived assuming that each star was in a binary system with the SMBH and neglecting the perturbations from other stars and gas. 
We notice that the maximum eccentricity $e_{\rm max}$ tends to increase with the semi-major axis: $e_{\rm max}$ goes from $\sim{}0.3$ for $a=0.14$ pc to $\sim{}0.4$ for $a=0.3$ pc. 
The simulated eccentricity range of the disk stars ($\langle{}e\rangle{}=0.29\pm{}0.04$) is in good agreement with the observed one for the clockwise disk surrounding SgrA$^\ast{}$ ($\langle{}e\rangle{}=0.36\pm{}0.06$; P06; Bartko et al. 2009). However, we stress that our simulations cover only the first $<0.5$ Myr since the formation of the gaseous disk, whereas the estimated age of the stars surrounding SgrA$^\ast{}$ is $\sim{}6$ Myr: the long-term dynamical evolution of the stellar disk might have affected the orbital properties of the stars. 

Fig.~\ref{fig:orbits} shows that, in our run~E, there is a gap in the distribution of semi-major axes between $a\sim{}0.30$ pc and $a\sim{}0.34$ pc. Furthermore, only a handful of stars  have $a>0.30$ pc. This depends on the structure of the gaseous disk, which is composed of many concentric annuli (as we said in Section~3.1). In particular, stars with $a<0.30$ pc in run~E come essentially from the same gaseous annulus, which has an outher radius $\approx{}0.30$ pc (see Fig.~\ref{fig:fig3}). Out of this annulus, there is a lower density region, surrounded by an outer high-density annulus (Fig.~\ref{fig:fig3}). The bunch of stars with $a\sim{}0.34$ pc formed in this outer annulus.

\begin{figure}
\center{{
\includegraphics[width=8.5cm]{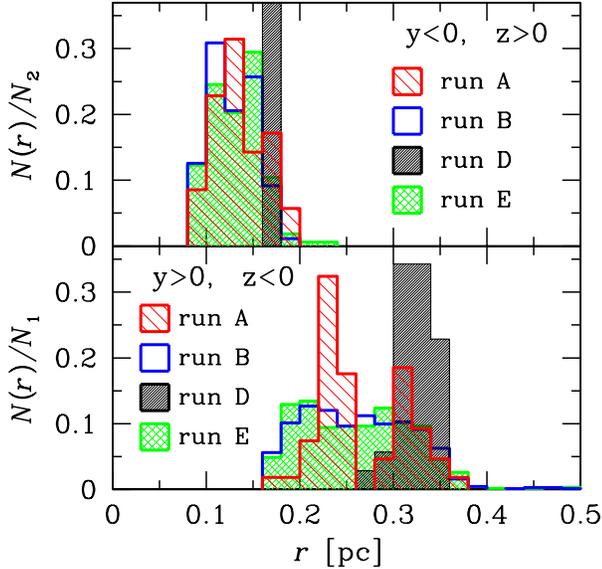} 
}}
\caption{\label{fig:fig5} 
Radial distribution of the star candidates at $t=4.8\times{}10^5$ yr. 
Top panel: stars close to the pericenter, i.e. with coordinates $y<0$ and $z>0$ (see Fig.~\ref{fig:fig3}). Bottom panel: stars close to the apocenter, i.e. with coordinates $y>0$ and $z<0$ (see Fig.~\ref{fig:fig3}).
$x-$axis: distance from the center $r$. $y-$axis: number of stars per radial bin $N(r)$, normalized to the total number of stars in a quadrant in each simulation ($N_1$ and $N_2$ for the quadrant with $y>0,\,{}z<0$ and for the quadrant with  $y<0,\,{}z>0$, respectively). Slightly hatched histogram (red on the web): run~A. Empty histogram (blue on the web): run~B. Heavily hatched histogram: run~D. Cross-hatched histogram (green on the web): run~E. Run~C is not shown, as no stars have formed at $t=4.8\times{}10^5$ yr.
}
\end{figure}

In runs~A, B, D and E, the first stars form in the outer parts of the ring, whereas stars form in the inner part (where the shear from the SMBH is stronger) only at later times. Furthermore, the clumps of gas particles tend to collapse while at the apocenter, where the distance from the SMBH is maximum. 
Figs.~\ref{fig:fig2}  and \ref{fig:fig3}  show that run~B and run~E are very similar.

Fig.~\ref{fig:fig5} shows the radial distribution of the stars that are closest to the pericenter, i.e. in the quadrant with $y<0$ and $z>0$ (top panel), and the radial distribution of the stars that are closest to the apocenter, i.e. in the quadrant with  $y>0$ and $z<0$ (bottom panel), at $t=4.8\times{}10^5$ yr. The top panel of this Figure shows that the stars at the pericenter are at a minimum distance from the SMBH $r_{\rm in}\sim{}0.10$ pc in run A, B and E, and $r_{\rm in}\sim{}0.16$ pc in run~D. The bottom panel of the same Figure indicates that the stars at the apocenter are at a maximum distance from the SMBH $r_{\rm out}\sim{}0.35-0.40$ pc in all the simulations.

\begin{figure}
\center{{
\includegraphics[width=8.5cm]{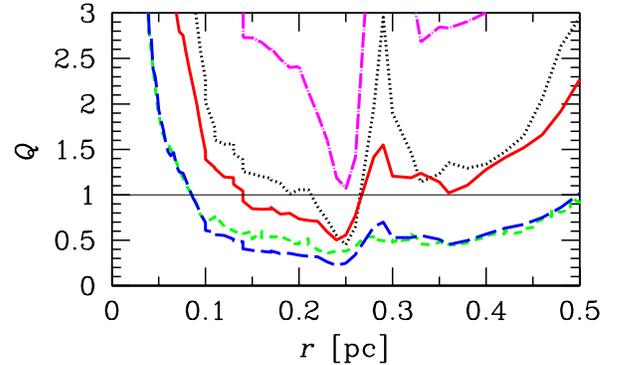} 
}}
\caption{\label{fig:fig6} 
Behavior of Toomre's $Q$ parameter slightly before the formation of the first stars (i.e. $t=4.8\times{}10^5$ yr for runs~A, B and E, $t=4.8\times{}10^5$ yr for runs~C and D). Solid line (red on the web): run~A. Long-dashed line (blue on the web): run~B. Dot-dashed line (magenta on the web): run~C. Dotted black line: run~D. Short-dashed line (green on the web): run~E.
Solid black thin line: $Q=1$.
}
\end{figure}

In the simulations, the existence of the inner radius  $r_{\rm in}$ is probably due to the fact that even the high central density of gas cannot counteract the Keplerian velocity at such small distances from the SMBH. In fact, Toomre's $Q$ parameter, defined as $Q(r)=\Omega{}(r)\,{}c_s(r)/(\pi{}\,{}G\,{}\Sigma{}(r))$ (where $\Omega{}(r)$ is the angular velocity, $c_s(r)$ the sound speed, $G$ the gravitational constant and 
$\Sigma{}(r)$ is the local surface density, Toomre 1964), is $Q(r)\gtrsim{}1$  at radii $r\lesssim{}0.15$, $0.1$, $0.2$ and $0.1$ pc for run~A, B, D and E, respectively (see Fig.~\ref{fig:fig6}). For higher values of $Q(r)$, the growth of gravitational instabilities in the disk is unlikely. The behavior of Toomre's $Q$ parameter also explains why gas clumps collapse predominantly at the apocenter of the orbit (see Fig.~\ref{fig:fig6}). Furthermore, we notice that, in Fig.~\ref{fig:fig6}, $Q(r)$ is always larger than 1 for run~C: this is consistent with the fact that no star candidates form in run~C. In addition, Fig.~\ref{fig:fig6} indicates that $r=0.5$ pc is a sort of cut-off for the SF in our simulations, as $Q(r)>1$ for $r>0.5$ pc in all the runs. On the other hand, this result strongly depends on the initial properties of the simulated cloud (density, mass, orbit, etc.). Moreover, $r=0.5$ pc should be interpreted as the tail end of a continuous distribution rather than a genuine cut-off.

The gas inside the inner radius remains in place till the end of the simulations. It would be interesting to assess whether this gas can be evaporated by the ultra-violet (UV) radiation and by the first supernovae of the young massive stars. Recently, Alexander et al. (2011) suggested that a fraction of the innermost gas (inside $\sim{}0.01$ pc) might have been used to power Sgr~A$^\ast{}$ in the past.

\begin{figure}
\center{{
\includegraphics[width=8.5cm]{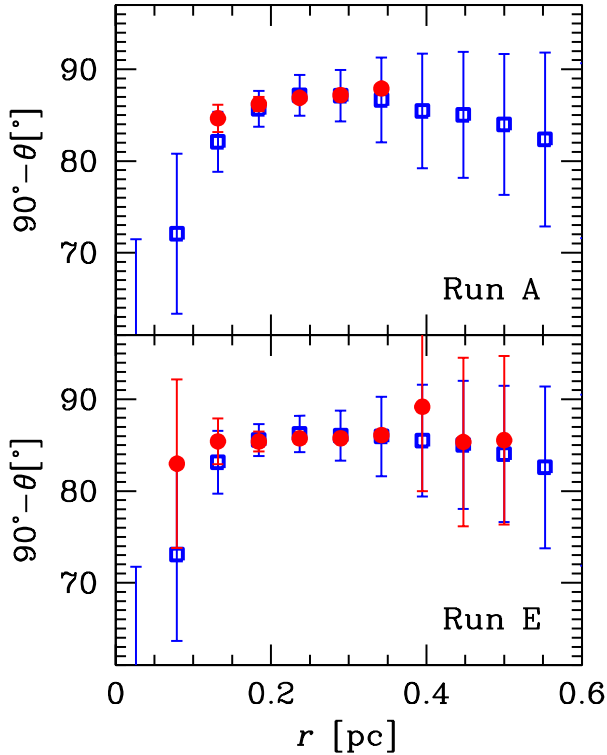} 
}}
\caption{\label{fig:fig7} 
Change of the angle $\theta{}$ (see definition in Section~3.2) as a function of the distance $r$ from the SMBH at $t=4.8\times{}10^5$ yr.
Filled circles (red on the web): binned values of $\theta{}$ for the star candidates. Open squares (blue on the web):  binned values of $\theta{}$ for the gas particles. Error bars show the standard deviation. Top panel: run~A. Bottom panel: run~E. We plot 90$^\circ{}-\theta{}$ for analogy with figure~13 of Bartko et al. (2009).
}
\end{figure}

We notice that the outer radius ($r_{\rm out}\sim{}0.4-0.5$ pc) and the thickness ($\zeta{}\sim{}0.05-0.08$)  of the stellar ring are slightly smaller than those of the gaseous disk (for which $r_{\rm out}\sim{}0.5$ pc  and $\zeta{}\sim{}0.1$ pc), but the stellar ring is inside the gaseous disk, has the same orientation, and the stellar orbits have the same eccentricity as those of the gaseous particles. Similarly to the gaseous disk, the stellar disk is irregular, quite warped and tilted. To quantify the importance of the warp and/or of the tilt, we study the radial change of the angular momentum orientation of stars and gas particles. In particular, we define the angle $\theta{}$ as
 $\cos{\theta{}}=\frac{L_{\rm max}}{L_{\rm tot}},$
 where $L_{\rm tot}$ is the modulus of the total angular momentum of a star or gas particle and $L_{\rm max}$ is the component of the angular momentum along the axis which maximizes the symmetry of the gas disk inside 1 pc radius (i.e., if the disk was perfectly axisymmetric, $L_{\rm max}$ would be along the symmetry axis of the disk). Fig.~\ref{fig:fig7} shows that there is a radial change by $\approx{}10^\circ{}$ in $\theta{}$  in both the gas and the stellar component, in all the considered runs (in the Figure, we show only runs~A and E, but the other three runs have very similar behavior). This indicates that both the gaseous and the stellar disks are slightly warped/tilted. Fig.~\ref{fig:fig7} cannot be directly compared with figure~13 of Bartko et al. (2009), as the angle $\theta{}$ shown therein is defined in a slightly different way (on the basis of the $\chi{}^2$ analysis described in Bartko et al. 2009). 
On the other hand, we notice that the trend of the angle in our paper and that shown in figure~13 of  Bartko et al. (2009) are similar.

The ring distribution of the star candidates is in good agreement with the observations (G03; P06). The outer radius $r_{\rm out}$ of the stellar ring is also in agreement with the observations (G03; P06), whereas the inner radius $r_{\rm in}$ is a factor of $\gtrsim{}2$ larger in the simulations than in the observations (which suggest $r_{in}\sim{}0.04$ pc). The fact that the stellar ring is irregular and that the stellar candidates have eccentric orbits with a large spread of eccentricities are also consistent with the observations (P06; Cuadra, Armitage \&{} Alexander 2008; Bartko et al. 2009).

\subsubsection{The mass function (MF)}
\begin{figure}
\center{{
\includegraphics[width=8.5cm]{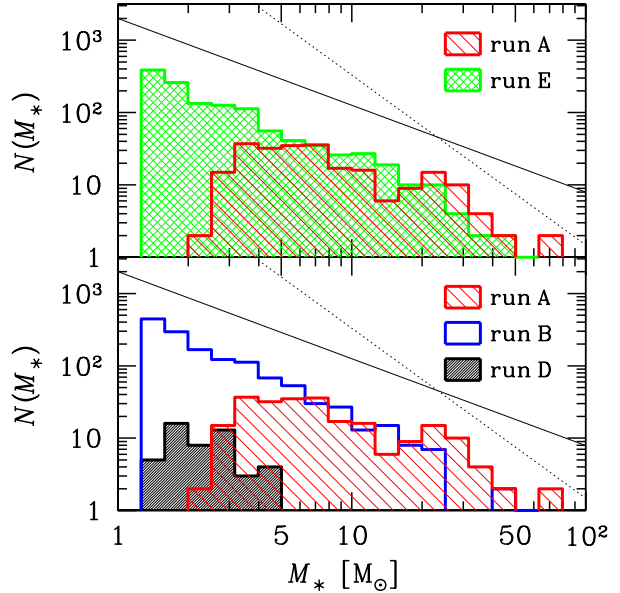} 
}}
\caption{\label{fig:fig8} 
Stellar MF in the simulations at $t=4.8\times{}10^5$ yr. $x-$axis: star mass $M_\ast{}$. $y-$axis: number of stars per mass bin $N(M_\ast{})$. 
In the bottom panel, slightly hatched histogram (red on the web): run~A; empty histogram (blue on the web): run~B; heavily hatched black histogram: run~D. In the top panel, slightly hatched histogram (red on the web): run~A; cross-hatched histogram (green on the web): run~E.
In both panels, solid (dotted) black thin line: MF $dN/dm\propto{}m^{-\alpha{}}$ with $\alpha{}=1.2$ ($\alpha{}=2.35$). Run~A is shown in both panels to allow easier comparison.
Run~C is not shown, as no stars have formed at $t=4.8\times{}10^5$ yr.         
}
\end{figure}

Fig.~\ref{fig:fig8} shows the MF of the stellar candidates for the different runs at $t=4.8\times{}10^5$ yr. Although we cannot exclude that the MF slightly evolves for $t>4.8\times{}10^5$ yr, Fig.~\ref{fig:fig8} is particularly interesting as it shows the differences among the various runs. Furthermore, the MF in Fig.~\ref{fig:fig8} slightly depends on our definition of star candidates and on the adopted threshold values. Thus, the normalization of the masses in Fig.~\ref{fig:fig8} might change  for different definitions of the proto-stars, but the differences among various runs are independent of such normalization.

\begin{figure}
\center{{
\includegraphics[width=8.5cm]{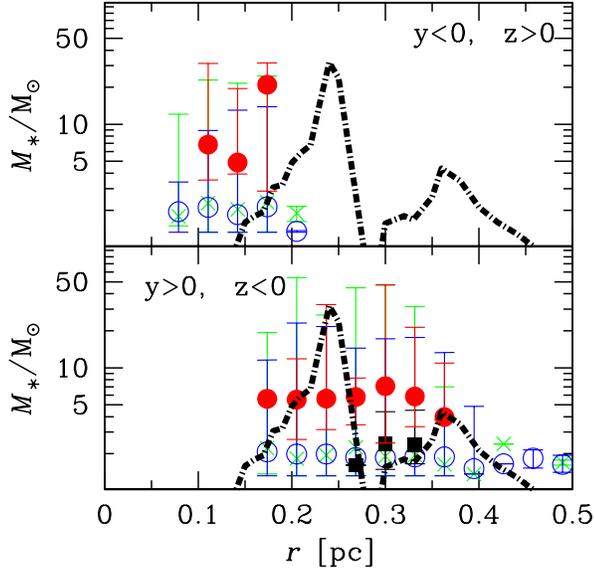} 
}}
\caption{\label{fig:fig9} 
Mass of the star candidates versus radial distance at $t=4.8\times{}10^5$ yr.  Filled circles (red on the web): run~A. Open circles (blue on the web): run~B. Black filled squares: run~D. Crosses (green on the web): run~E.
 Run~C is not shown, as no stars have formed at $t=4.8\times{}10^5$ yr.
 The points correspond to the median stellar mass per each bin, whereas the upper (lower) error bars show the maximum (minimum) stellar mass.
Dot-dashed line: Toomre mass for run~A.
Top panel: stars close to the pericenter, i.e. with coordinates $y<0$ and $z>0$ (see Fig.~\ref{fig:fig3}). Bottom panel: stars close to the apocenter, i.e. with coordinates $y>0$ and $z<0$ (see Fig.~\ref{fig:fig3}).
}
\end{figure}

Fig.~\ref{fig:fig8} indicates that the MF is heavier in run~A than in run~B,~D and E. 
 Fig.~\ref{fig:fig8} also shows that the number of star candidates formed in runs~B (1363) and E (1252) is much larger than in runs~A (238) and ~D (49).
However, the total mass of stars in run~A,  $M_{\rm tot}=2.35\times{}10^3\,{}{\rm M}_\odot{}$, is $\sim{}20$ times larger than that in run~D ($M_{\rm tot}=1.18\times{}10^2\,{}{\rm M}_\odot{}$) and only a factor of $\lesssim{}2$ lower than that in run~B  ($M_{\rm tot}=3.98\times{}10^3\,{}{\rm M}_\odot{}$) and in run~E ($M_{\rm tot}=4.33\times{}10^3\,{}{\rm M}_\odot{}$), since the MF in run~A is heavier than in the other runs. We note that the total mass of stars in runs~A,~B and E matches the value predicted by the observations ($\approx{}5\times{}10^3\,{}{\rm M}_\odot{}$), although we have already stressed the $caveats$ of the normalization of our simulated MFs. 

 \begin{deluxetable}{lll}
 \tabletypesize{\tiny}
 \tablewidth{0pt}
 \tablecaption{Least square fit parameters.}
\tablehead{\colhead{Run}
& \colhead{$\alpha{}$}
& \colhead{$\beta$}}
\startdata
A       & 0.5$\pm{}0.2$ & 1.6$\pm{}0.3$ \\
B       & 1.56$\pm{}0.04$ & 2.85$\pm{}0.04$ \\
D       & 0.7$\pm{}$0.7 & 1.1$\pm{}0.3$ \\
E       & 1.50$\pm{}$0.06 & 2.81$\pm{}0.07$ 
\enddata
 \end{deluxetable}


Table~2 shows the least square fit of $\log{}N=-\alpha{}\,{}\log{}M_\ast{}+\beta{}$ for runs A, B and E\footnote{In Table~2, we do not report run~C, because no stars form in this run. In run~D the number of stellar candidates is too low to distinguish among different MFs: we report the results of run~D in Table~2, but the uncertainty on the fit parameters is huge.}. The best fitting MF is heavier than the Salpeter MF ($\alpha=2.35$) for all the considered runs.
The MF of run~E (with radiative cooling) is quite similar to that of run~B (isothermal), whereas the MF of run~A is significantly heavier than in both run~B and E. The MF of run~E (which differs from run~B only for the inclusion of radiative cooling) deviates from that of run~B mainly because of a high mass tail, that is not present in run~B: 19 (10) stars form in run~E (B) with mass $>20$ M$_\odot{}$.
The fact that runs~A, B and E have a MF heavier than the Salpeter MF is important, as the observations of young stars in the GC indicate that they are relatively massive.  We notice that the best-fitting slope of the MF in run~A ($\alpha{}=0.5$) is in fair agreement with the most recent observations of the stellar disk(s) in the GC (see, e.g., $\alpha{}=0.45\pm{}0.3$ from fig.~3 of Bartko et al. 2010).

Intuitively, the reason why the MF in run~A is heavier than in runs~B and E (and the MF in all the performed runs is heavier than the Salpeter MF) is connected with the gas temperature. In fact, the Jeans mass (Jeans 1919) scales with ${\rm T}_{\rm MC}^{3/2}$. This implies that there should be a factor of $\approx{}30$ difference in the typical mass of stars formed in a MC at ${\rm T}_{\rm MC}=10$ K (which is a typical temperature for MCs, in absence of strong UV background) with respect to stars formed in a MC at ${\rm T}_{\rm MC}=100$ K (which is a typical temperature in the neighborhoods of the GC, Nagai et al. 2007).
Similarly, there should be a factor of $\sim{}10$ difference between the typical mass of stars in run~A and that of stars in runs~B and E. Actually, the average star mass in run~A is $\langle{}M_\ast{}\rangle{}=10\,{}{\rm M}_\odot{}$, only a factor of $\sim{}3-4$ larger than the average star mass in run~B  ($2.9\,{}{\rm M}_\odot{}$). The discrepancy between the difference expected from the Jeans instability and the difference obtained from the simulations is probably due to the resolution limit of the simulations. In fact, we cannot resolve stars with mass $M_\ast{}<1.28$ M$_\odot{}$.



 


Fig.~\ref{fig:fig9} shows the radial dependence of the star mass ($M_\ast{}$) in the simulations at $t=4.8\times{}10^5$ yr. The top panel of Fig.~\ref{fig:fig9} shows the radial distribution of mass only for the stars closest to the pericenter. All these stars are at $0.08\le{}r/{\rm pc}\le{}0.2$. However, this distribution is not relevant to understand their formation, as most of stars form while at the apocenter (see explanation in the previous section). Therefore, stars that are at the pericenter at $t=4.8\times{}10^5$ yr have likely formed at larger radii. The bottom panel is more interesting, as it shows the distribution of stars that are at the apocenter at $t=4.8\times{}10^5$ yr. In fact, such distribution approximately mirrors that of the stellar birthplaces.
In all the simulations there is no significant correlation between the star mass and the radial distance from the SMBH. In run~D, stars form at $0.25\le{}r/{\rm pc}\le{}0.35$, whereas in runs~A, B and E there is a wider spread ($0.15\le{}r/{\rm pc}\le{}0.4$). In runs~B and E there are also a few stars at larger radii than $0.4$ pc. 

\begin{figure}
\center{{
\includegraphics[width=8.5cm]{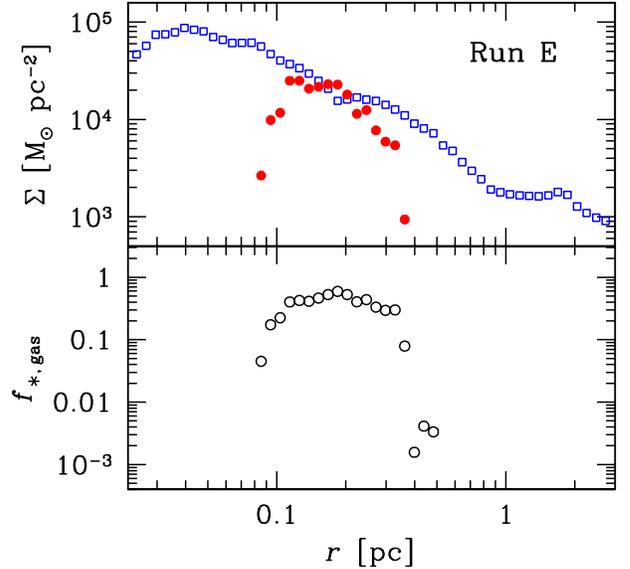} 
}}
\caption{\label{fig:surface} 
Top panel: mass surface density of gas (open squares, blue on the web) and stars (filled circles, red on the web) in run~E at $t=4.8\times{}10^5$ yr. Bottom panel: $f_{\ast,\,{}{\rm gas}}$ (i.e. SF efficiency) as a function of radius in run~E at $t=4.8\times{}10^5$ yr.}
\end{figure}

In Fig.~\ref{fig:fig9} we also compare the radial mass distribution with the estimates from Toomre's most unstable wavelength. Toomre's most unstable wavelength, defined as $\lambda{}_{\rm T}\equiv{}2\,{}\pi{}^2\,{}G\,{}\Sigma{}/\kappa{}^2$ (where $\Sigma{}$ is the local surface density and $\kappa{}$ is the epicyclic frequency, Binney \& Tremaine 1987) in the case of a thin gaseous disk, is the wavelength at which instability first appears, when $Q$ drops below unity in a differentially rotating disk. The dot-dashed line in Fig.~\ref{fig:fig9} shows the Toomre's mass, defined as $M_{\rm T}\equiv{}\frac{4\,{}\pi{}}{3}\,{}\rho{}\,{}\lambda{}_{\rm T}^3$, in the case of run~A. The distribution of stars in runs A, B and E is in partial agreement with the predictions from Toomre's mass. The differences are likely due to the fact that Toomre's mass does not account for the eccentricity of the disk, for further gas accretion and for changes in the local surface density. Moreover Toomre's mass assumes that the disk is in equilibrium.

\subsubsection{The SF efficiency}
 We can quantify the efficiency of SF. The surface density $\Sigma{}$ of stars (filled circles) and of gas (open squares) are both shown in the top panel of Fig.~\ref{fig:surface}: the mass density of stars is comparable with that of gas only at $r=0.1-0.2$ pc and falls abruptly at smaller and larger radii. The SF efficiency $f_{\ast,\,{}{\rm gas}}$ (defined as $f_{\ast,\,{}{\rm gas}}=M_\ast{}/(M_\ast{}+M_{\rm gas})$, where $M_{\rm gas}$ and $M_{\ast{}}$ are the gas mass and the stellar mass, respectively) reaches a maximum of $\sim{}0.6$ at $r=0.18$ pc in run~E, and stays above 0.1 in the $r=0.09-0.4$ pc range. Out of this range, it drops very quickly. A possible issue may be the removal and/or evaporation of the remaining gas surrounding the massive stars. On the other hand, our code does not include radiative feedback (e.g., UV emission) from the newly formed stars and cannot describe the evolution of the gas after the formation of the proto-stars. For the same reason,  we cannot quantify the importance of further gas accretion onto the formed stars, over longer timescales than $t=4.8\times{}10^5$ yr. Furthermore, the estimated age for the young stars surrounding SgrA$^\ast{}$ is about 6 Myr: at this epoch, the supernovae from the most massive stars are expected to have already taken place.

\subsection{Properties of gas}
\begin{figure}
\center{{
\includegraphics[width=8.5cm]{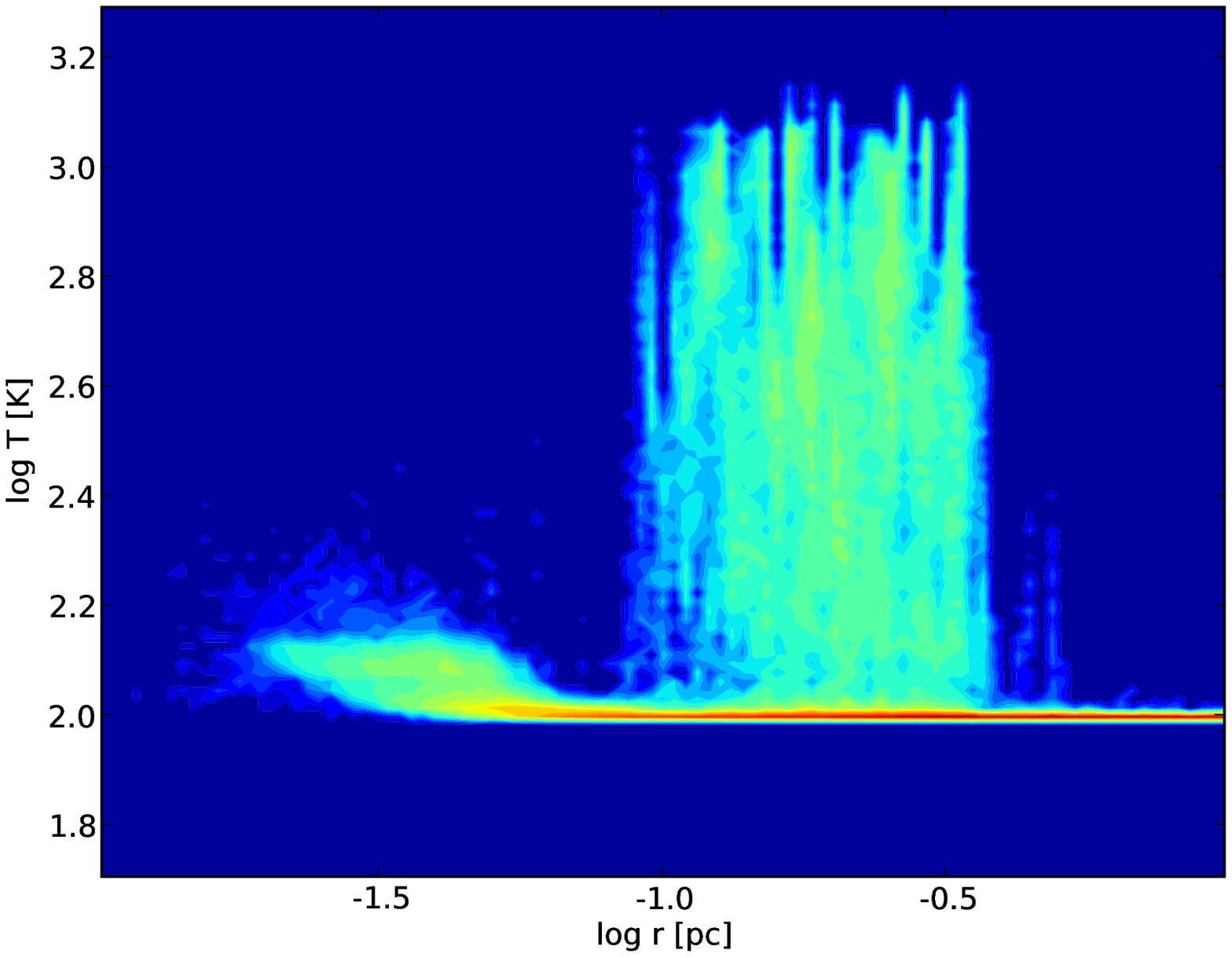}  
\includegraphics[width=8.5cm]{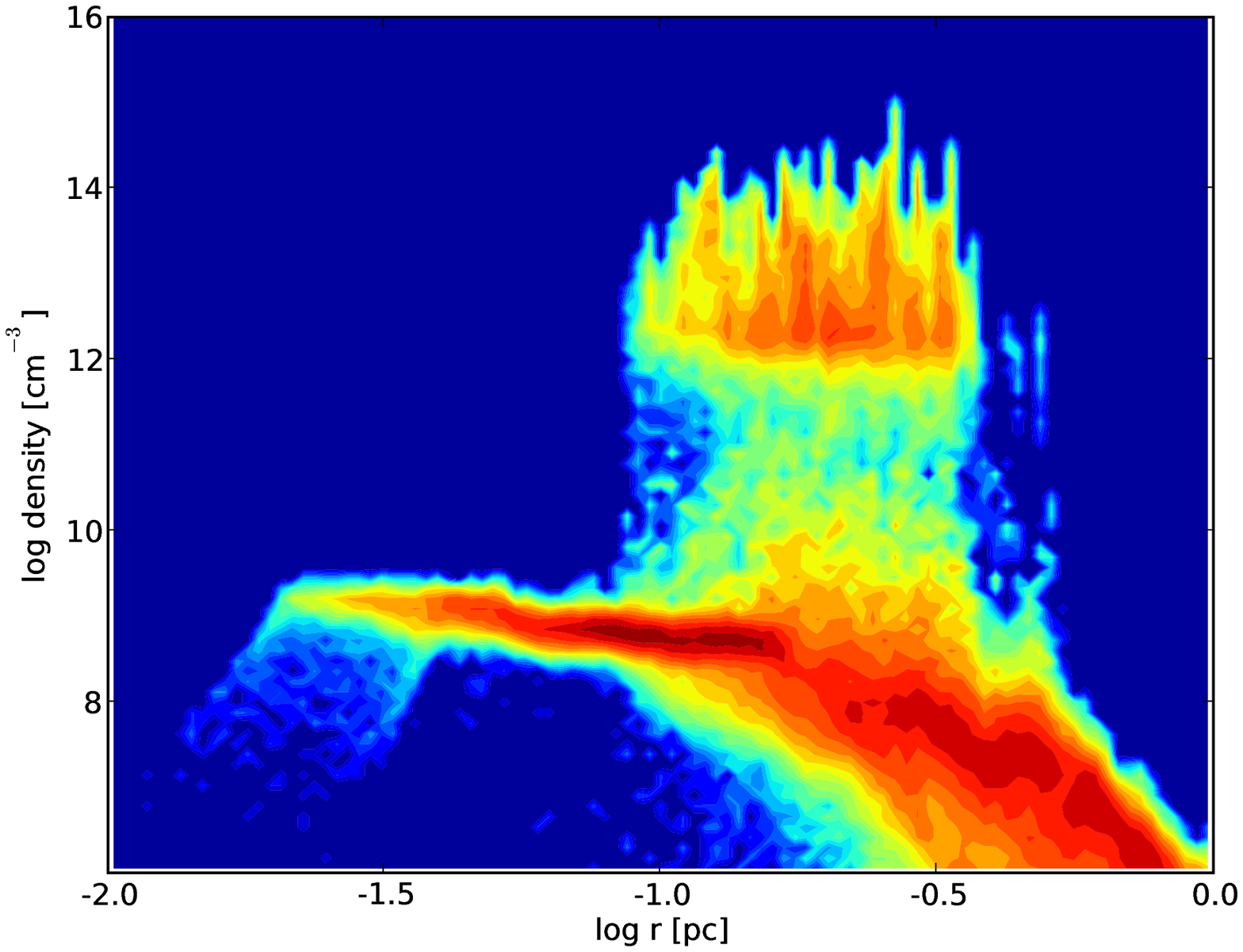}  
\includegraphics[width=8.5cm]{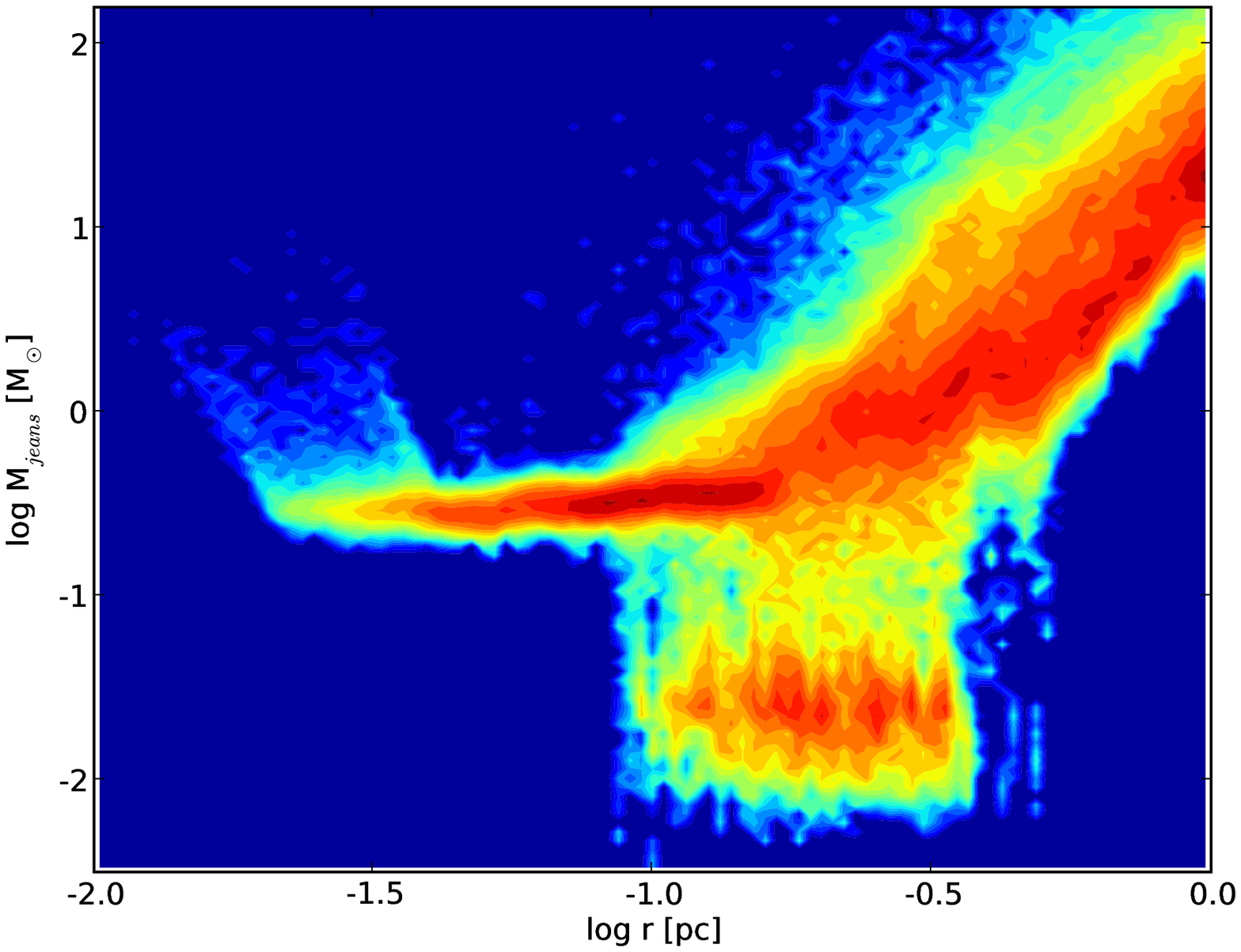}  
}}
\caption{\label{fig:fig11}
Top, central and bottom panel: contour plot of gas temperature, density and Jeans mass, respectively, versus radial distance from the SMBH, in run~E, at $t=4.8\times{}10^5$ yr. 
}
\end{figure}

Run~E includes accurate recipes for gas cooling (see Section~2). Therefore, we focus on run~E to analyze the local properties of gas in the disk. Fig.~\ref{fig:fig11} shows the radial behavior of gas temperature (top panel),  density (central) and Jeans mass (bottom). 
The temperature plot shows that most of gas is close to the temperature floor ${\rm T}_{\rm irr}=100$ K. Gas temperature increases only very close to the SMBH, and at  $r\sim{}0.1-0.4$ pc inside the collapsed clumps (i.e. the star candidates), because of the much higher density and of the consequent compressional heating.

The plot of the density (central panel of Fig.~\ref{fig:fig11}) is particularly interesting because it shows three (almost distinct) regimes. For $r\lesssim{}0.1$ pc, the local density in  the disk is high ($\approx{}10^{9}$ cm$^{-3}$) and almost constant, with a very slow rise towards the SMBH. For $r>0.1$ pc, the local density in most of the disk drops quite fast ($\rho{}\propto{}r^{-\phi{}}$, where the slope is $\phi{}\approx{}4$). Independently of this global behavior, a bunch of gas particles at $r\sim{}0.1-0.4$ pc reach very high densities ($>10^{12}$ cm$^{-3}$): this represents the phase of star candidates.


Finally, Jeans mass (bottom panel of Fig.~\ref{fig:fig11}) confirms that we actually resolve most of the star candidates in run~E, at least for $r\geq{}0.2$ pc. In fact, the expected Jeans mass is always $\sim{}0.3$ M$_\odot{}$ for $r\leq{}0.2$ pc. This means that we cannot resolve the small stars that might form at $r\leq{}0.2$ pc, as our mass resolution is 1.28 M$_\odot$. For larger radii the Jeans mass increases, as a consequence of the fast decrease in density, and becomes larger than our mass resolution. We stress that the subsample of gas particles with Jeans mass $<<0.1$ M$_\odot{}$ in Fig.~\ref{fig:fig11} is spurious: these gas particles are within the already collapsed clumps and the Jeans mass is no longer a valid estimator for them, as their density is already decoupled from the background density.
We do not plot the behavior of  density and Jeans mass in run~B, because it is very similar to run~E.

\begin{figure}
\center{{
\includegraphics[width=8.5cm]{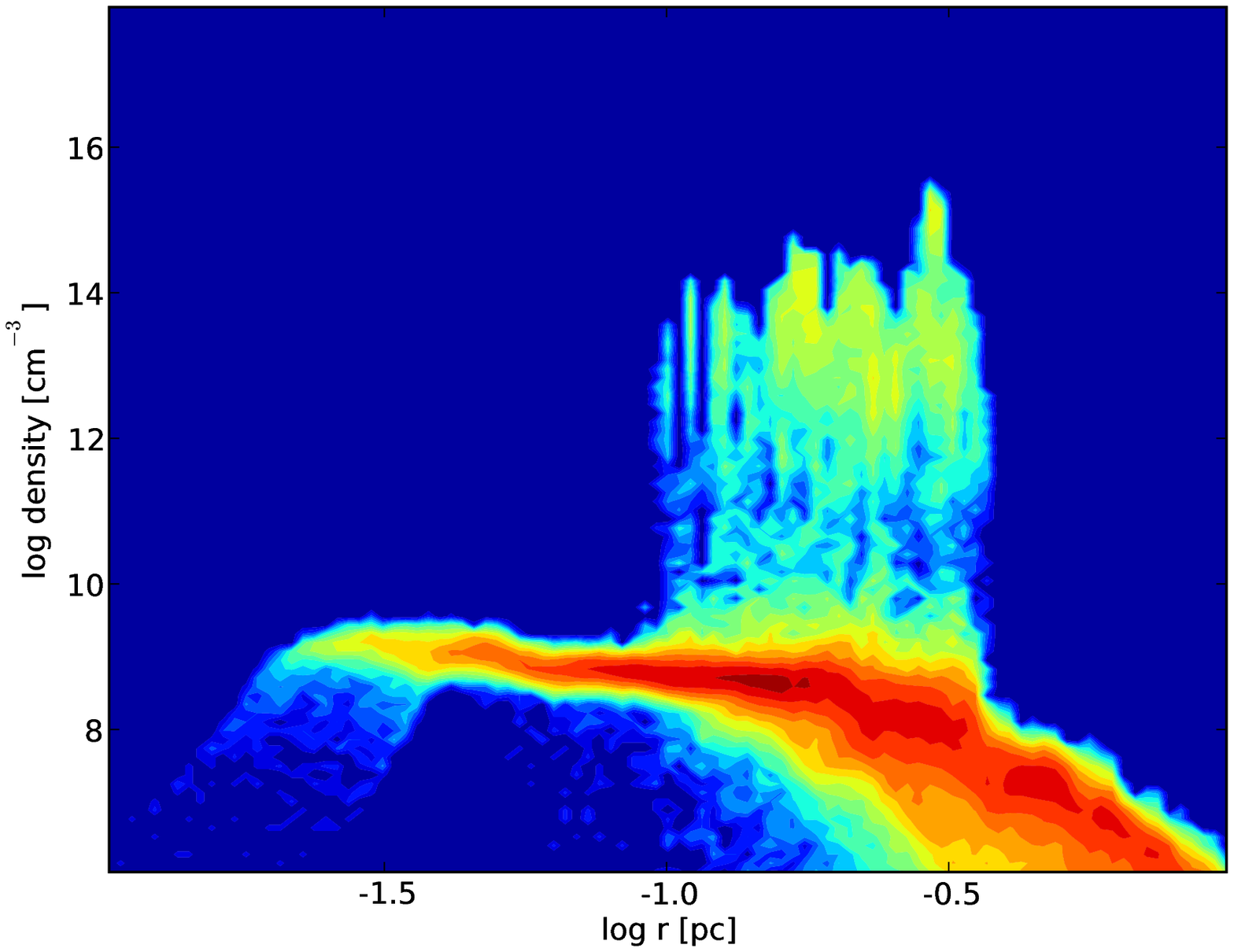} 
\includegraphics[width=8.5cm]{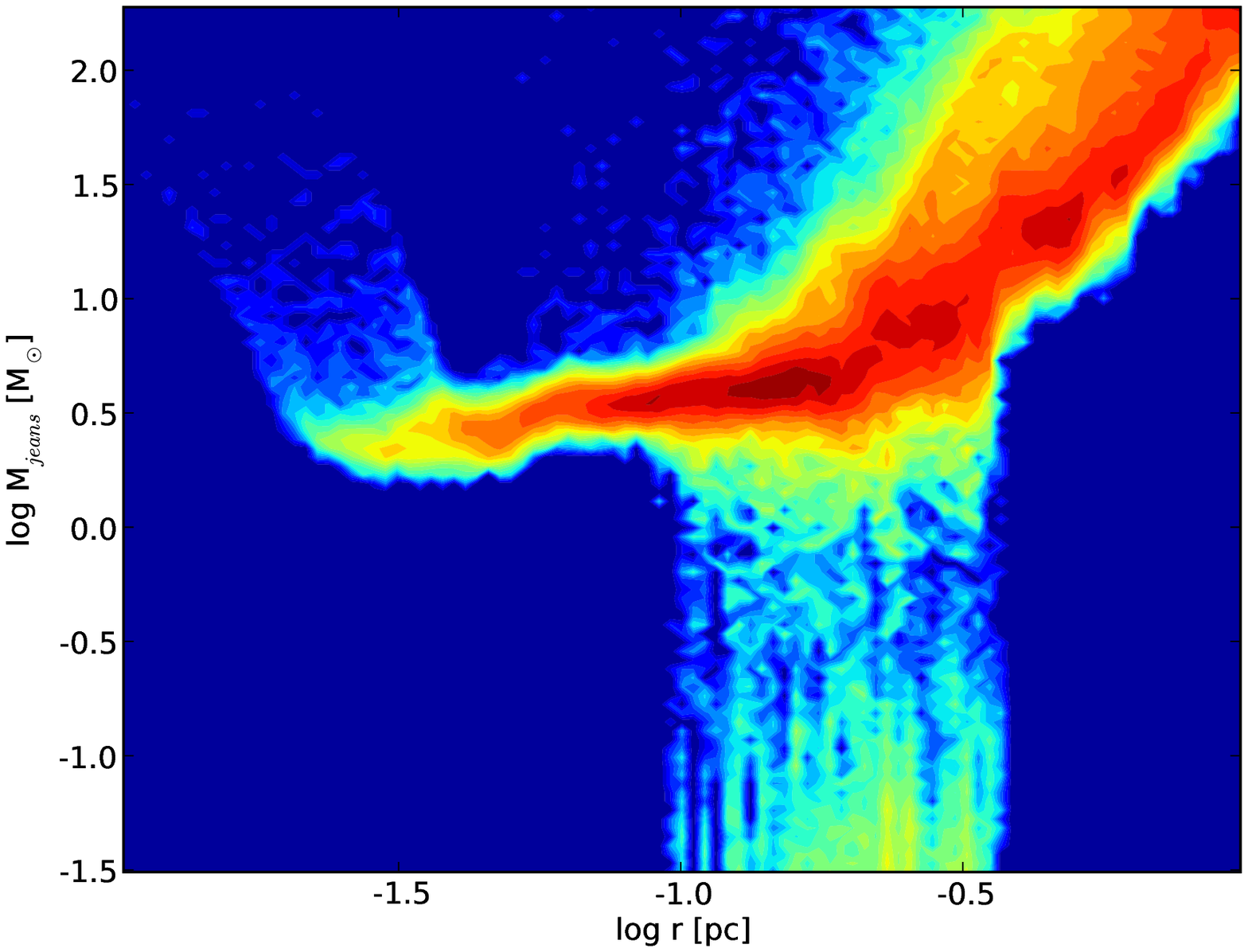} 
}}
\caption{\label{fig:fig12}
Top (bottom) panel: contour plot of gas density (Jeans mass) versus radial distance from the SMBH, in run~A, at $t=4.8\times{}10^5$ yr. 
}
\end{figure}

 Fig.~\ref{fig:fig12} shows the behavior of density and Jeans mass for run~A. The behavior of density (top panel) is not very different from that of run~E, apart from the fact that the number of collapsed clumps (star candidates) is lower. Instead, the Jeans masses are a factor of $\sim{}5-10$ higher.  This means that the Jeans mass is $\approx{}2-6$ M$_\odot{}$ for $r\lesssim{}0.3$ pc.
We notice that the minimum mass of star candidates in run~A from Fig.~\ref{fig:fig8} is $2-3$ M$_\odot{}$, that is we resolve most of the star candidates in run~A.

\subsection{Comparison with previous papers}
In this Section, we briefly compare our simulations and our main results with those reported in previous papers.  In particular, we consider the four papers whose approach is closest to ours, i.e. M08, BR08, HN09 and A11. We first compare the methods adopted in these papers, and then their results.

\subsubsection{Different methods}
 All these papers, as well as ours, simulate the infall of a molecular cloud towards the GC, its disruption from the SMBH, and the formation of a dense gaseous disk. The adopted softening  and mass resolution are comparable in all these simulations. The simulations in M08, HN09, A11 and this paper were run for similar evolutionary times ($\approx{}2.5-6\times{}10^5$ yr), whereas the simulations in BR08 were run only for $\approx{}3-5\times{}10^4$ yr. This difference is mainly due to the orbit of the simulated cloud(s). 

M08, BR08 and this paper adopt models of gas cloud that are turbulently supported, while HN09 and A11 consider a simplified model of spherical and homogeneous cloud. The radius of the cloud is quite different in these studies: M08 and the current paper adopt a radius of 15 pc, consistent with the radii of Galactic molecular clouds; A11 consider clouds with initial radius of 3.5 pc; BR08 consider very compact clouds with a radius of 1 pc. In the case of HN09 the involved clouds are small (with a radius of $0.17-0.2$ pc), but this cannot be compared directly with the other papers, as HN09 simulate a collision between the two clouds, which changes deeply the properties of the involved gas.

M08, HN09 and this paper include a cuspy rigid potential to account for the other stars close to the GC (see, e.g., G03), whereas BR08 and A11 do not include any underlying rigid potentials. The inclusion of a spherical potential contributes to stabilize the disk (delaying or preventing local fragmentation), although the mass of the SMBH dominates any other gravitational contributions in the region of interest ($\lesssim{}0.4$ pc).

The simulations in M08 are isothermal, with T$_{\rm MC}=10$~K, respectively. Such temperature is likely too low, if compared to the background radiation field in the GC (see e.g. Nagai et al. 2007). 
In BR08 the simulations include an approximate radiative transfer formalism, with compressional heating balanced by cooling rates derived from estimated optical depths. In HN09, the simulations include a very simplified model of cooling (see their equation~1), and the initial temperature of the cloud is very low (T$_{\rm MC}=20$ K). HN09 do not specify whether there is a temperature floor in their simulations and do not provide a quantitative description of temperature evolution (see their appendix for a discussion about cooling).
In A11 and in this paper, the simulations include different thermodynamical treatments for the gas, including both isothermal and radiative cooling cases (even an adiabatic run in A11). The floor temperature for the simulations with radiative cooling is set to be 50 K in A11 and 100 K in both BR08 and this paper. For a further analysis of the differences between our radiative transfer formalism and that used in BR08, see Appendix~B.

A11 stop their simulation before fragmentation takes place in the disk, whereas the other considered papers study the formation of star candidates in the disk. M08, BR08 and HN09 adopt the sink particle technique, to model SF. Only in this paper we follow the fragmentation of the disk, although our resolution is border-line for the lower star masses ($M_\ast{}\lesssim{}2$ M$_\odot{}$). The attempt of directly following the fragmentation, without using the sink particles, is the main improvement of this paper with respect to the previous ones.
\subsubsection{Comparison of the results}
All the considered papers indicate that a dense gaseous disk may form around the GC as a consequence of the infall and of the disruption of a massive molecular cloud.
If the impact parameter is non-zero, the resulting gaseous disk will be eccentric, consistently with the observations of the stellar orbits in Sgr~A$^\ast{}$. The eccentricity and the size of the final gaseous disk are similar in all the considered papers. 

The main differences among various simulations refer to the formation of star candidates. M08 adopted a value of the critical density for converting gaseous particles into sink particles ($=10^6$ cm$^{-3}$) that is too low if compared to the tidal density induced by the presence of the SMBH inside the disk. Such assumption generally leads to the formation of sink particles with larger masses than those we expect for cold ($\sim{}10$ K) clouds. 
HN09 assume a more robust critical density, weighted with the local tidal density ($\rho{}_{\rm crit}=4.1\times{}10^{11}$ cm$^{-3}+50\,{}\rho{}_{\rm tid}$, where $\mu{}=2.46$ is assumed and $\rho{}_{\rm tid}$ is the tidal density). This choice of the critical density for converting gaseous particles into sink particles is very close to our criterion for SF (see Section 3.2). On the other hand, the MF in HN09 is quite low-mass (especially in their runs S3 and S4), because of the approximations in the cooling recipe and because of the absence of opacity prescriptions.

BR08 adopt a very conservative value of the critical density for converting gaseous particles into sink particles ($=10^{14}$ M$_\odot{}$ pc$^{-3}=1.6\times{}10^{15}$ cm$^{-3}$, assuming $\mu{}=2.46$), well above the critical tidal density. Therefore, their MF is consistent with that predicted by the Jeans mass for the local density and temperature of the clouds.

Similarly, the MFs derived in this paper (Fig.~\ref{fig:fig8}) are consistent with the predictions calculated from Jeans mass and  Toomre instability. Furthermore, the fact that we do not use sink particles prevents any possible bias connected with that method. The selection of star candidates on the basis of  $\rho{}_{\rm th}$ (see Section 3.2) is shown to be robust by the analysis of Fig.~\ref{fig:fig11}. Confirming the results by BR08, we show that the MF in the stellar ring may be top-heavy only if the parent cloud is sufficiently massive ($\gtrsim{}10^5$ M$_\odot{}$) and if sufficiently high temperatures ($\gtrsim{}100$ K) are reached during the SF. 

On the other hand, the differences between our run~B (massive isothermal cloud with T$_{\rm MC}=100$ K) and our run~E (massive cloud with radiative cooling and T$_{\rm irr}=100$ K) are quite small. Furthermore, we do not observe in our run~E the formation of the very massive stars ($>60$ M$_\odot{}$) that were found in the massive cloud simulated by BR08. The MF derived from run~E is consistent with a single power-law with index $\alpha{}\sim{}1.5$, whereas the MF by BR08 is clearly bimodal, showing two distinct stellar populations (see fig.~4 by BR08). The very massive stars in BR08 are all formed at $r\sim0.02$ pc, where high temperatures and Jeans masses are reached (as it can be seen from their figs. S1, S2 and S3). In our simulations, star candidates never form at $r<0.05$ pc, because the shear from the SMBH prevents local collapse. Instead, in the simulation by BR08, the formation of sink particles is very efficient at $r\sim0.02$ pc (see their fig.~2). We point out that the mass of the cloud, the mass resolution of gas particles, the softening and smoothing lengths and the temperature floor in the two simulations are very similar. The reasons of the difference are likely the following. First, the orbit of the two clouds are different, leading to different distributions of position and angular momentum around the SMBH. We start integrating when the cloud is 25 pc far from the GC, whereas  BR08 assume a much smaller initial distance (3 pc). Furthermore, we adopt a cloud radius $d=15$ pc, whereas BR08 assume that the cloud radius is only $d=1$ pc (for approximately the same cloud mass). This implies that the density of gas streams in the disrupting cloud of BR08 is significantly higher than in our simulations. This is evident also from Figs.~\ref{fig:fig11} and ~\ref{fig:fig12}, where only few gas clumps reach the density $\rho{}=1.6\times{}10^{15}$ cm$^{-3}$, that is the density threshold for the conversion of gas into sinks in BR08.
Furthermore, our implementation of the opacity and that used by BR08 have some differences. In particular, Stamatellos et al. (2007, from which BR08 take their formalism) have a deep opacity gap at temperature T$_{\rm MC}\lesssim{}1000$ K, when the refractory elements start to vaporize  (see Appendix~B). In addition, Bell \&{} Lin (1994) Rosseland opacity (adopted by BR08) is too low at T$_{\rm MC}\sim{}1500-1800$ K (see, e.g., Alexander \&{} Ferguson 1994), that is the temperature at which most of massive stars form in BR08. This might justify a heavier MF in BR08 than in our paper.
 In addition, we do not use the sink particle method (apart from the SMBH), whereas BR08 do. In particular, BR08 adopt a very robust minimum density ($\sim{}2\times{}10^{15}$ cm$^{-3}$) to create new sink particles. Instead, we assume that star particles form where the local density of gas is $>\rho{}_{\rm th}=2\times{}10^{12}$ cm$^{-3}$. Our density threshold $\rho{}_{\rm th}$ is three orders of magnitude smaller than the threshold by BR08, but is well motivated, because clumps of particles with density above $\rho{}_{\rm th}$ are gravitationally bound and have a thermal behavior completely distinct from the rest of the disk (see our Fig~\ref{fig:fig11} and Section 3.3). It may be that the higher density threshold required by BR08, together with the sink recipes, introduces a selection bias in the formation of stars and 
 suppresses the formation of low-mass stars (see, e.g. the discussion in Appendix~A).
We stress that all very massive stars in BR08 form very close ($<0.02$ pc) to the GC, where massive stars have not been observed in the Milky Way (the observed ring of young stars having an inner radius of $\sim{}0.04$ pc). From this point of view, our run~E seems to match better the observational properties of the Milky Way.

\section{Summary and conclusions}
We simulate the infall of a molecular cloud toward SgrA$^\ast{}$. We ran five different clouds, with various masses ($4.3\times{}10^4$, $1.3\times{}10^5$ M$_\odot{}$), initial temperatures (100, 500 K) and thermodynamics (isothermal or radiative cooling). In all the simulations, the cloud is disrupted by the tidal forces of the SMBH and spirals towards it. In less than $10^5$ yr, more than one tenth of the gas in the parent cloud ends up into a dense, eccentric and distorted disk around the SMBH, with a small outer radius ($\sim{}0.5$ pc). Locally, the surface density of the gaseous disk may overcome the tidal shear from the SMBH and fragmentation may take place. We define as `star candidates' those gas clumps that decouple from the background gas and collapse. This is the first attempt to trace the fragmentation of the disk, without adopting the sink particle technique.

Star candidates form in four of the five simulations within the first $4.8\times{}10^5$ yr. Only run~C, in which the cloud is small ($4.3\times{}10^4$ M$_\odot{}$) and warm (500 K), does not host star candidates at $t=4.8\times{}10^5$ yr, but stars might form at later times.
 
In the remaining four simulations, star candidates are distributed in a thin ring at a distance of $\sim{}0.1-0.4$ pc from the SMBH. They have eccentric orbits ($0.2\le{}e\le{}0.4$), with average eccentricity $\langle{}e\rangle{}=0.29\pm{}0.04$, in agreement with the properties of the young massive stars observed around Sgr~A$^\ast{}$ ($\langle{}e\rangle{}=0.36\pm{}0.06$, Bartko et al. 2009).

In the three runs (A, B and E) where the cloud is more massive ($1.3\times{}10^5$ M$_\odot{}$), the total mass of star candidates ($2-4\times{}10^3$ M$_\odot{}$) is consistent with the observations (P06). Instead, the total mass of stars is only $\sim{}10^2$ M$_\odot{}$ in run~D, where the cloud is smaller ($4.3\times{}10^4$ M$_\odot{}$). Therefore, our simulations suggest that a rather massive molecular cloud ($\gtrsim{}10^5$ M$_\odot{}$) is required, in order to form the stars observed around the GC. 

Furthermore, the MF of stellar candidates is very top-heavy (fitted by a single power-law with $\alpha{}<1$) only in run~A, for which T$_{\rm MC}=500$ K. In runs B and E (for which T$_{\rm MC}=100$ K) the MF (fitted by a single power-law with $\alpha{}\sim{}1.5$) is still heavier than the Salpeter MF.
Run~E, which includes radiative cooling, differs from run~B (isothermal) only for the presence of a high mass tail in the MF. 

This is consistent with the predictions from Jeans instability. Therefore, high gas temperatures (T$_{\rm MC}\gtrsim{}100$ K) should be reached during the SF, in order to generate a top-heavy MF, such as the one observed in the GC. Such high temperatures are not unrealistic in the neighborhood of Sgr~A$^\ast{}$.
We stress that, in addition to the total mass of the cloud, the density of the original cloud can strongly affect the stellar MF. This is likely the main difference between our simulations and those by BR08, together with the usage of sink particles.


A possible limit of our simulations is the mass resolution: we barely resolve $\sim{}1-2$ M$_\odot{}$ stars. Higher resolution simulations are important, to improve our understanding of SF around SMBHs. 
Furthermore, in our five runs the cloud has always the same orbital parameters. It would be interesting to study the influence of the cloud orbit onto the formation of the gaseous disk and of the stellar ring. This issue will be also addressed in a forthcoming paper. 
In this paper, we consider only a single molecular cloud. HN09 find that the collision between two gas clouds is naturally able to explain the existence of more than one stellar disk. Therefore, the scenario of the cloud collision deserves further investigations, with accurate cooling and opacity prescriptions.

A number of other open questions deserve further studies. For example, this paper focuses on the formation of the stars in the disk, but the long-term dynamical evolution of these stars cannot be studied with the same numerical approach (and we remind that the observed massive stars in the GC have an estimate age of $6$ Myr). Various papers addressed this issue before, with somewhat different results (see, e.g.,  Alexander et al. 2007; Cuadra et al. 2008; L\"ockman et al. 2010). 
 
 A possibly related issue is the origin of the so-called `S-stars' (e.g., Sch\"odel et al. 2002; Ghez et al. 2003; Bartko et al. 2010). These are B stars that do not belong to the disk(s), as have isotropic spatial distribution and higher eccentricities. They are predominantly located in the inner 0.05 pc, but recent studies (Bartko et al. 2010) show that their distribution extends further out. Our simulations cannot explain at the same time young disk stars and isotropic S-stars, because they are not consistent with the same stellar population. On the other hand, it is likely that the disruption of two different clouds at different times can account for both populations (e.g., HN09; Perets \&{} Gualandris 2010). In this scenario, the S-stars are the survivors of a previous disk and their orbits have been randomized by the influence of a perturber (e.g., Perets et al. 2007; Yu et al. 2007; Perets et al. 2008; Gualandris \&{} Merritt 2009; Perets \&{} Gualandris 2010), or by a Kozai-like resonance between two disks (L\"ockmann et al. 2008, 2009), or  by spiral density wave migration in the gaseous disk (Griv 2010). We propose that even the evaporation of the parent gaseous disk (due, e.g., to the local UV and X-ray emission) might randomize the orbits of the stars, if the SF efficiency was sufficiently low. 
On the other hand, we cannot exclude on the basis of our simulations a completely different origin for the young disk(s) and for the S-stars (e.g., Alexander \&{} Livio 2004; Fujii et al. 2010; Perets \&{} Gualandris 2010; Baruteau, Cuadra \&{} Lin 2011; Madigan et al. 2011).



\begin{figure}
\center{{
\includegraphics[width=8.5cm]{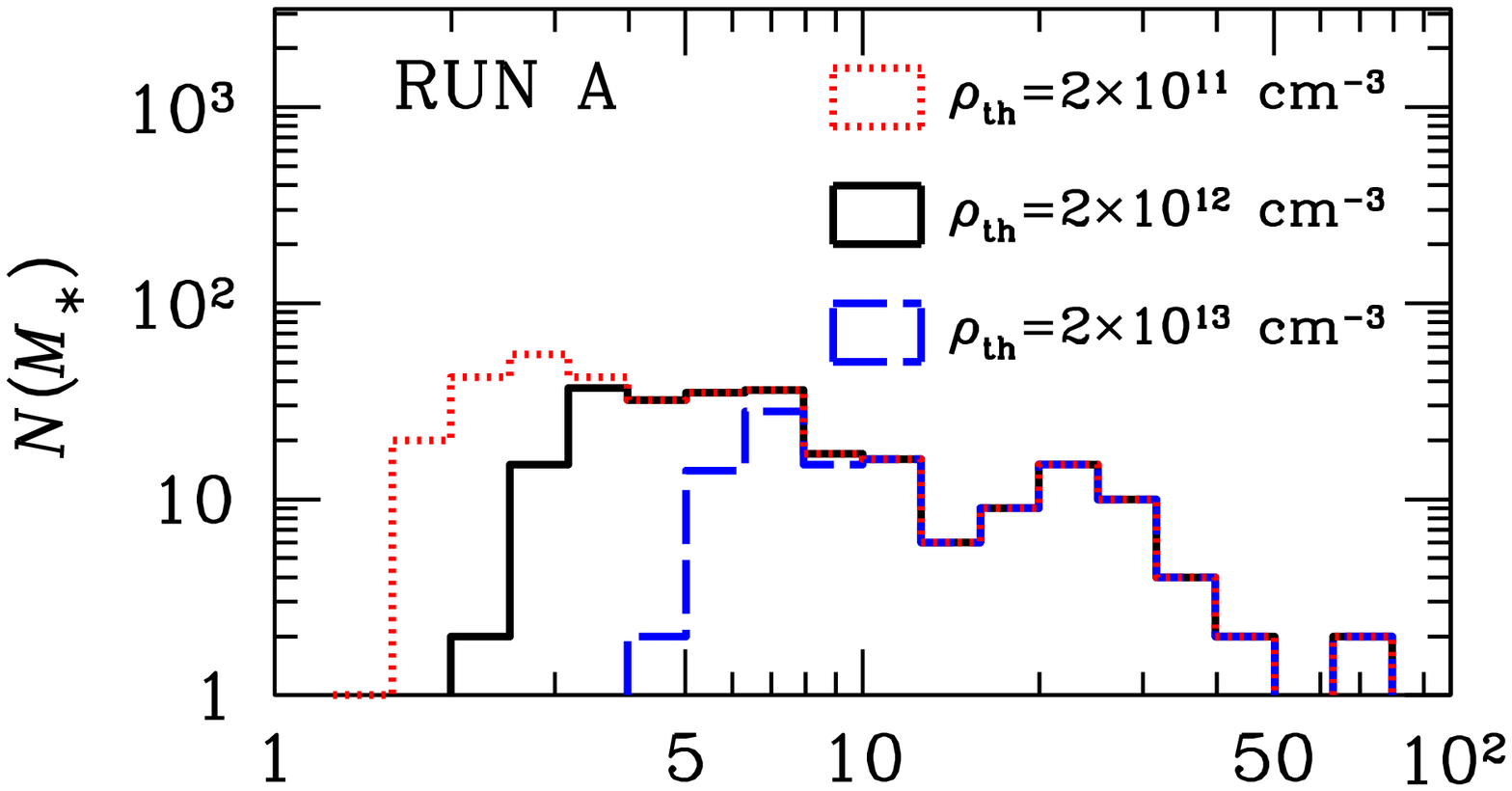} 
\includegraphics[width=8.5cm]{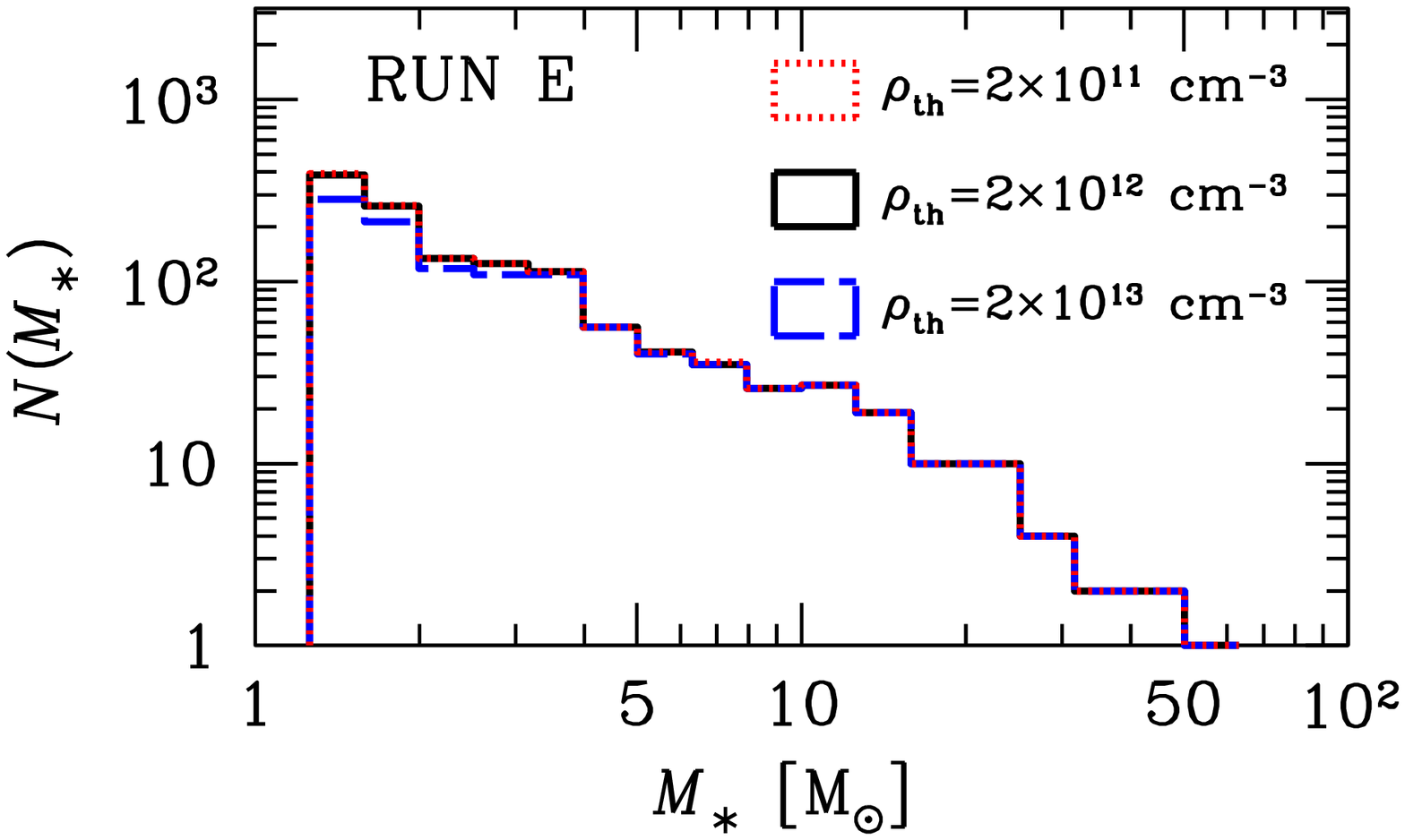} 
}}
\caption{\label{fig:figA1} 
Stellar MF in run~A (top panel) and in run~E (bottom panel) at $t=4.8\times{}10^5$ yr. $x-$axis: star mass $M_\ast{}$. $y-$axis: number of stars per mass bin $N(M_\ast{})$. Dotted line (red on the web): the adopted threshold density is $\rho{}_{\rm th}=2\times{}10^{11}$ cm$^{-3}$. Black solid line: the adopted threshold density is $\rho{}_{\rm th}=2\times{}10^{12}$ cm$^{-3}$ (the same as in Fig.~\ref{fig:fig8}). Dashed line (blue on the web): the adopted threshold density is $\rho{}_{\rm th}=2\times{}10^{13}$ cm$^{-3}$. 
}
\end{figure}
\begin{figure}
\center{{
\includegraphics[width=8.5cm]{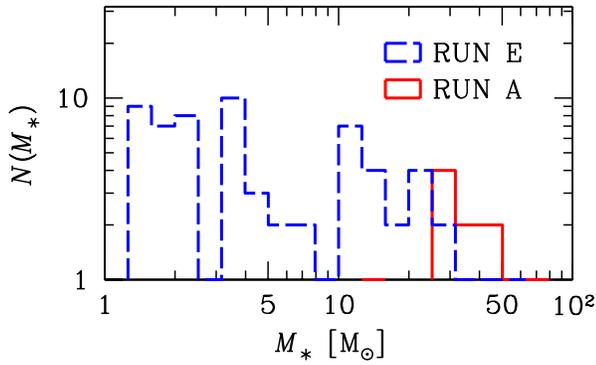} 
}}
\caption{\label{fig:figA2} 
Stellar MF in run~A (solid line, red on the web) and in run~E (dashed line, blue on the web) at $t=4.8\times{}10^5$ yr, adopting $\rho{}_{\rm th}=2\times{}10^{15}$ cm$^{-3}$. $x-$axis: star mass $M_\ast{}$. $y-$axis: number of stars per mass bin $N(M_\ast{})$.}
\end{figure}
\begin{figure}
\center{{
\includegraphics[width=8.5cm]{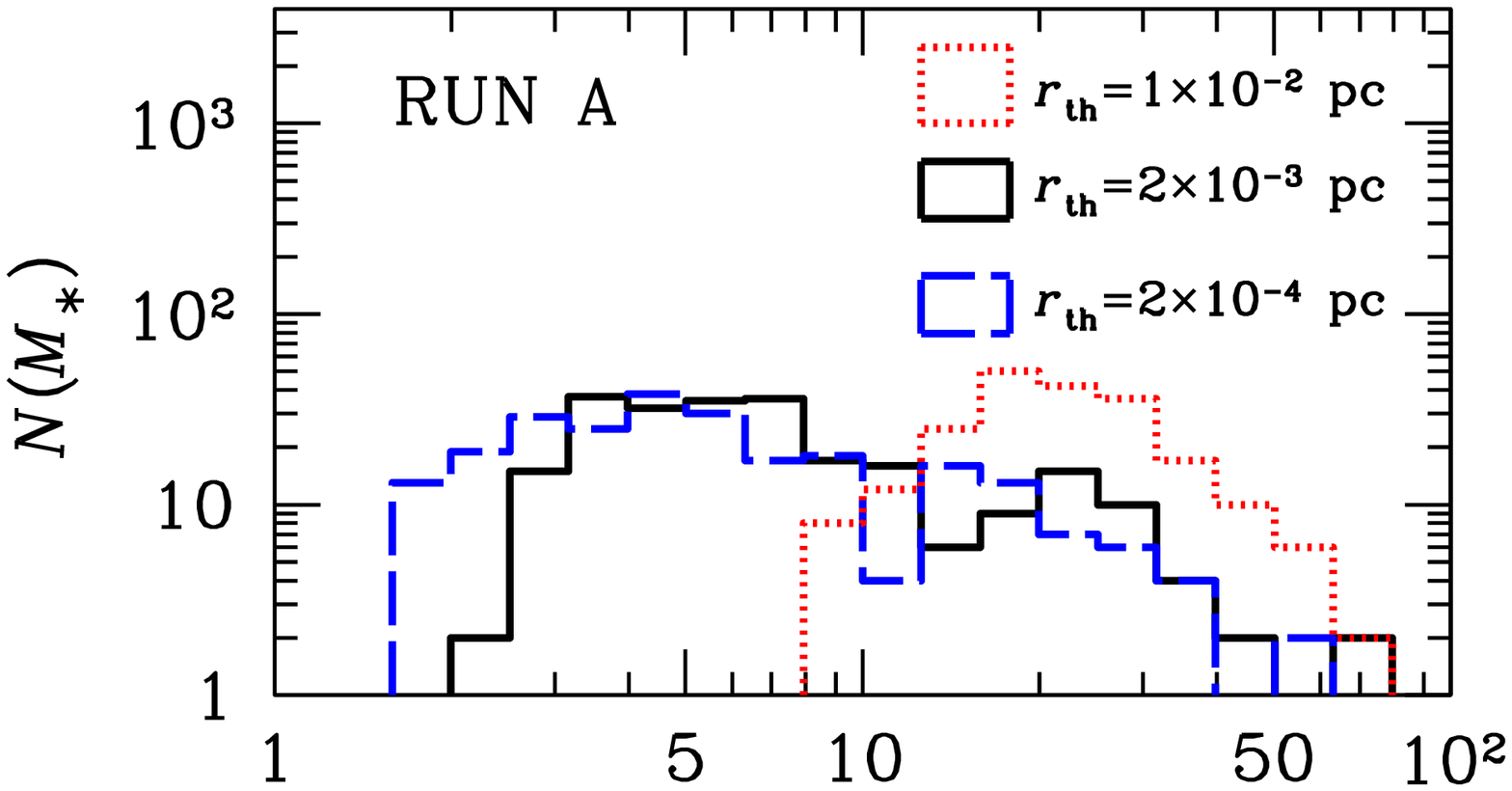} 
\includegraphics[width=8.5cm]{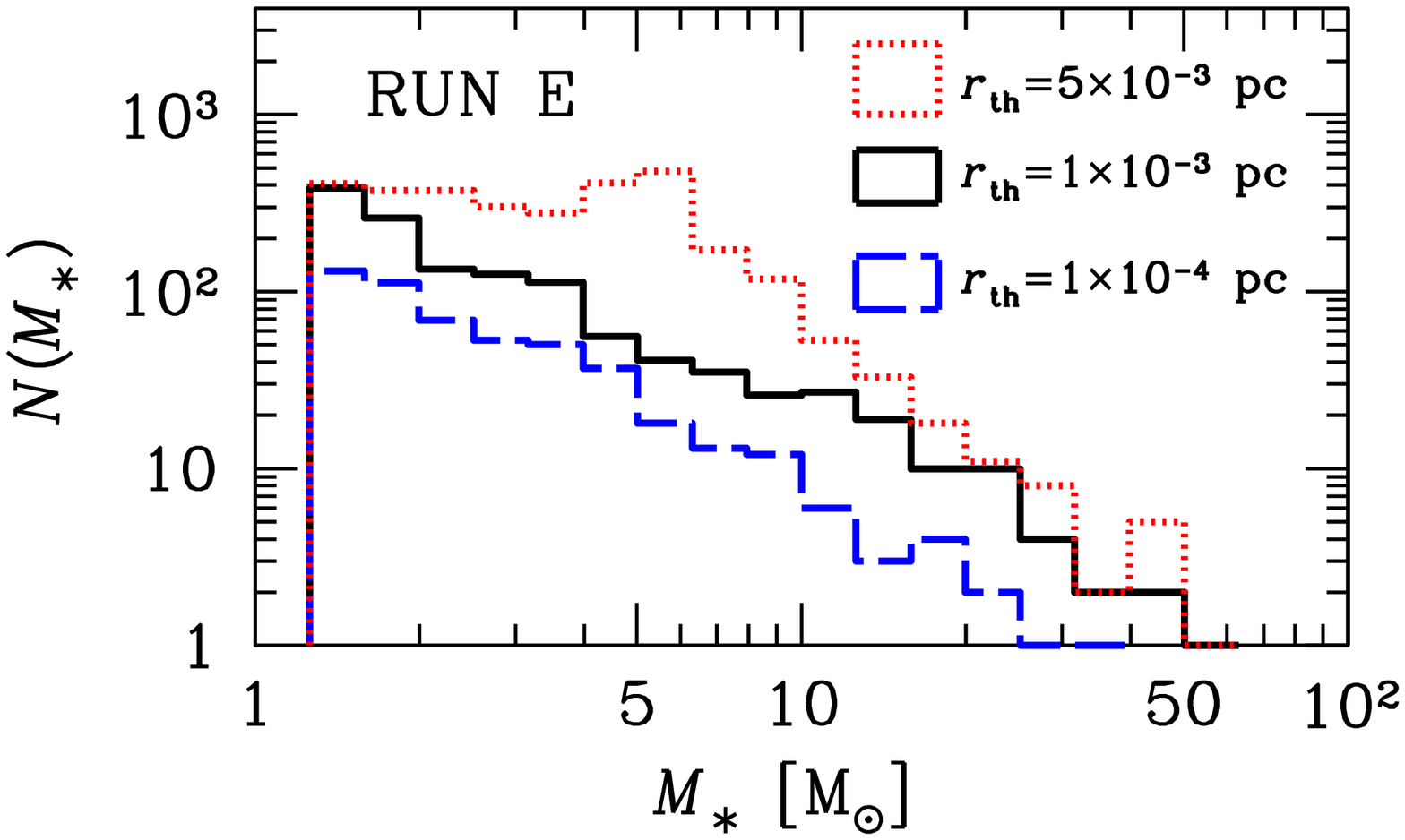} 
}}
\caption{\label{fig:figA3} 
Stellar MF in run~A (top panel) and in run~E (bottom panel) at $t=4.8\times{}10^5$ yr. $x-$axis: star mass $M_\ast{}$. $y-$axis: number of stars per mass bin $N(M_\ast{})$. Dotted line (red on the web): the adopted threshold radius is $r_{\rm th}=1\times{}10^{-2}$ pc  ($r_{\rm th}=5\times{}10^{-3}$ pc) in the top (bottom) panel. 
Black solid line: the adopted threshold radius is $r_{\rm th}=2\times{}10^{-3}$ pc ($r_{\rm th}=1\times{}10^{-3}$ pc) in the top (bottom) panel.
Dashed line (blue on the web): the adopted threshold radius is $r_{\rm th}=2\times{}10^{-4}$ pc ($r_{\rm th}=1\times{}10^{-4}$ pc) in the top (bottom) panel. 
}
\end{figure}

\begin{figure}
\center{{
\includegraphics[width=7.5cm]{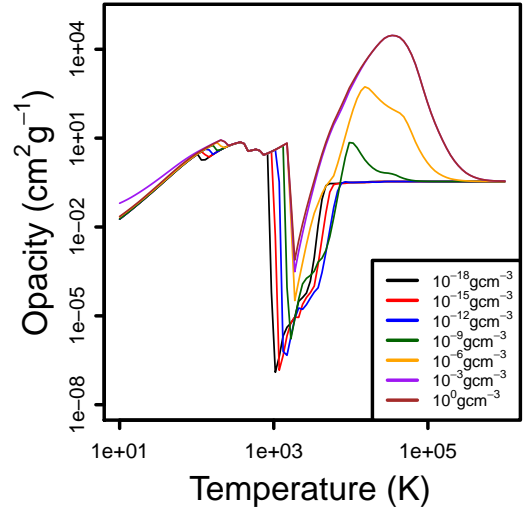} 
}}
\caption{\label{fig:figB1}
Opacity as a function of temperature from our code. Isopycnic curves are plot from $\rho= 10^{-18}$, to $\rho{} = 1$ g cm$^{-3}$ and spaced by a factor of $10^3$ (bottom to top).
}
\end{figure}


\section*{Acknowledgments}
We thank the anonymous referee for providing constructive comments that improved the paper, and D. Semenov for enlightening discussions about opacity. We also thank A. Gualandris, B. Moore, E.~D'Onghia, E. Ripamonti and P.~Englmaier for useful discussions, and acknowledge J. Stadel and D. Potter for technical support. The simulations were run onto the CRAY XT3 cluster at the Swiss National Supercomputing Center in Lugano, onto the $zbox3$ cluster at the Institute for Theoretical Physics of the University of Zurich and onto the `lagrange' cluster at the Consorzio Interuniversitario Lombardo per L'Elaborazione Automatica (CILEA). MM, TH and LM   acknowledge support from the Swiss National Science Foundation.


{}
\begin{appendix}
\section{How critical is the choice of the threshold density?}
We adopt a threshold density
 $\rho{}_{\rm th}=2\times{}10^{12}$ cm$^{-3}$ (assuming molecular weight $\mu{}=2.46$) and a threshold radius $r_\ast{}\le{}r_{\rm th}=2.2\times{}10^{-3}\,{}({\rm T}_{\rm MC}/500)^{1/2}$~pc, in order to identify star candidates. Such choices are reasonable, as  the adopted $r_{\rm th}$  is of the same order of magnitude as the softening length (therefore, we cannot consider smaller radii) and the adopted $\rho{}_{\rm th}$ is at least $10^3$ times larger than the local tidal density (therefore, the gravitational torque by the SMBH cannot stop the collapse of these clumps).

Slightly different values of $r_{\rm th}$ and of $\rho{}_{\rm th}$ do not change significantly our results. Fig.~\ref{fig:figA1} shows that the MF of star candidates in run A is affected only for stellar masses $M_\ast{}\lesssim{}6\,{}M_\odot{}$, if we change $\rho{}_{\rm th}$ from $2\times{}10^{11}$ cm$^{-3}$ to $2\times{}10^{13}$ cm$^{-3}$. In particular, lower mass stars are suppressed by larger choices of $\rho{}_{\rm th}$. In the case of run~E, where the gas is initially at a lower temperature (T$_{\rm MC}=100$ K), the MF is substantially unchanged, even when changing $\rho{}_{\rm th}$ by one order of magnitude.

Interestingly, BR08 adopt a significantly higher threshold value ($\rho_{\rm sink}\sim{}2\times{}10^{15}$ cm$^{-3}$), in order to convert gas clumps into sink particles. If we use the same density as $\rho{}_{\rm th}$ in our simulations, we obtain the results shown in Fig.~\ref{fig:figA2}. A much lower number of star candidates is derived, by assuming this threshold, and most of the low-mass star candidates are suppressed.


Fig.~\ref{fig:figA3} shows how different choices of the threshold radius $r_{\rm th}$ can affect the MF. Smaller values of $r_{\rm th}$ (down to about one tenth of the fiducial value) do not change significantly the MF. Similarly, the MF is not significantly affected by larger values of $r_{\rm th}$, up to $2-5$ times the fiducial value. The choice of larger values of $r_{\rm th}$ is more critical: the MF becomes significantly heavier in all the considered runs. However, there are no physical reasons to adopt such large values of $r_{\rm th}$.

In conclusion, the choice of $r_{\rm th}$ and of $\rho{}_{\rm th}$ is crucial for the final MF. We adopt physically well motivated values for both $r_{\rm th}$ and $\rho{}_{\rm th}$. Slightly different choices of $r_{\rm th}$ and $\rho{}_{\rm th}$ do not affect significantly the MF. Significantly different choices of $r_{\rm th}$ and $\rho{}_{\rm th}$ (e.g., $r_{\rm th}\gtrsim{}10^{-2}$ pc and $\rho_{\rm th}\gtrsim{}10^{15}$ cm$^{-3}$) change both the normalization and the slope of the MF. However, in the current study, we are mainly interested in the relative differences between various runs, that are almost not affected by the choice of $r_{\rm th}$ and of $\rho{}_{\rm th}$.

\section{Different treatments of opacity}
In run~E, we use the same radiative cooling algorithm as that described in Boley (2009) and in Boley et al. (2010). D'Alessio et al. (2001) Rosseland and Planck opacities are used in our code. 

BR08 use the formalism described in Stamatellos et al. (2007). They compute Rosseland opacities using Bell \&{} Lin (1994) recipes. 

The primary differences are: (i) Stamatellos et al. (2007) fold the gas potential (but not the star potential) and density into an (effective) pseudo density, and we do not. (ii) Stamatellos et al. (2007) gas energy equation includes a whole range of source terms related to dissociation and ionization of H, H$_2$ and He. Our gas energy equation does not. (iii) Bell \&{} Lin (1994) opacities are (orders of magnitude) too low at T$_{\rm MC}\sim{}1500-1800$ K, because they truncate the water opacities at too short a wavelength (D. Semenov, private communication; see also Alexander \&{} Ferguson 1994).

The fact that we do not include source terms related to dissociation of H is an issue for our model, but only for temperatures above $\sim{}2000$ K, that are basically never reached in our run~E (see Fig.~\ref{fig:fig11}). Instead, the fact that Bell \&{} Lin (1994) opacities (adopted by BR08) are too low at T$_{\rm MC}\sim{}1500-1800$ K may have serious consequences on the MF, as most of massive stars in BR08 form in that range of temperatures (see their figure S1). Too low opacities might artificially boost fragmentation, inducing spurious SF at temperatures (T$_{\rm MC}\sim{}1500$~K) where the Jeans mass is high.

In Fig.~\ref{fig:figB1}, we plot our opacity versus temperature, along isopycnals from $\rho= 10^{-18}$, to $\rho{} = 1$ g cm$^{-3}$ (bottom to top).
This does not totally match figure 5 of Stamatellos et. al (2007), that have a very deep opacity gap at T$_{\rm MC}~ 1000$~K, when the refractory elements start to vaporize. This might partially explain why BR08 MF is top-heavier than ours.

\end{appendix}
\end{document}